\documentclass[a4paper,fleqn,usenatbib]{mnras}
\usepackage{newtxtext,newtxmath}
\usepackage[T1]{fontenc}
\usepackage{ae,aecompl}
\usepackage{graphicx}	
\defcitealias{sfd}{SFD98}
\defcitealias{ccm89}{CCM89}
\defcitealias{schlafly2016}{SMS16}
\defcitealias{davenport2014}{DIB14}
\defcitealias{av}{G12}
\defcitealias{g17}{G17}
\defcitealias{lallement2014}{LVV14}
\defcitealias{lallement2019}{LVV19}
\defcitealias{green2015}{GSF15}
\defcitealias{green2018}{GSF18}
\defcitealias{green2019}{GSF19}
\defcitealias{arenou}{AGG92}
\defcitealias{drimmel}{DCL03}
\defcitealias{wang2019}{WC19}
\defcitealias{ruiz2018}{RBA18}
\defcitealias{2015ApJ...798...88M}{MF15} 

\title[{\it Gaia} DR2 giants in the Galactic dust]
{{\it Gaia} DR2 giants in the Galactic dust -- 
I. Reddening across the whole dust layer and some properties of the giant clump.}

\author[G. A. Gontcharov and A. V. Mosenkov]{
George A. Gontcharov,$^{1,2}$\thanks{E-mail: george.gontcharov@tdtu.edu.vn}
and Aleksandr V. Mosenkov$^{3}$
\\
$^{1}$Department for Management of Science and Technology Development, Ton Duc Thang University, Ho Chi Minh City, Vietnam\\
$^{2}$Faculty of Applied Sciences, Ton Duc Thang University, Ho Chi Minh City, Vietnam\\
$^{3}$Central Astronomical Observatory, Russian Academy of Sciences, 65/1 Pulkovskoye chaussee, St. Petersburg, 196140 Russia
}

\date{Accepted 2020 September 07. Received 2020 September 03; in original form 2020 April 15}

\pubyear{2020}

\begin{document}
\label{firstpage}
\pagerange{\pageref{firstpage}--\pageref{lastpage}}
\maketitle

\begin{abstract}
We consider a complete sample of 101\,810 giants with {\it Gaia} Data Realease 2 (DR2) parallaxes $\varpi$ within the red clump domain of the 
Hertzsprung--Russell diagram in the space cylinder with a radius of 700 pc around the Sun and a height of $|Z|=1800$ pc.
We use the {\it Gaia} DR2 $G_\mathrm{BP}$, $G_\mathrm{RP}$ and {\it Wide-field Infrared Survey Explorer (WISE)} $W3$ photometry.
We describe the spatial variations of the modes of the observables 
$G_\mathrm{BP}-G_\mathrm{RP}$, 
$G_\mathrm{RP}-W3$, 
$G_\mathrm{BP}+5+5\,\log_{10}\varpi$,
$G_\mathrm{RP}+5+5\,\log_{10}\varpi$, and 
$W3+5+5\,\log_{10}\varpi$
by extinction and reddening in combination with linear vertical gradients of the intrinsic colours and absolute magnitudes of the red giant clump.
The derived clump median absolute magnitude in $W3$ agrees with its recent literature estimates.
The clump median intrinsic colours and absolute magnitudes in $G_\mathrm{BP}$ and $G_\mathrm{RP}$ are derived for the first time at a precision level of 0.01 mag.
We confirm the reliability of the derived clump absolute magnitudes, intrinsic colours, and their vertical gradients by comparing them with the theoretical predictions 
from the PAdova and TRieste Stellar Evolution Code, MESA Isochrones and Stellar Tracks and Bag of Stellar Tracks and Isochrones isochrones.
This leads us to the median age and [Fe/H] of the clump within $|Z|<1.7$ kpc from the Galactic mid-plane as $(2.3\pm0.5)+(3.2\pm1.6)\,|Z|$ Gyr and 
$(-0.08\pm0.08)-(0.16\pm0.07)\,|Z|$ dex, respectively, where $Z$ is expressed in kpc. These results agree with recent empirical and theoretical estimates.
Moreover, all the models give similar age--metallicity relations by use of our results in the optical range.
The derived extinctions and reddenings across the whole dust half-layer below or above the Sun 
converge to the reddening $E(B-V)=0.06$ mag by use of the most reliable extinction laws.
\end{abstract}


\section{Introduction}
\label{intro}

The properties of the Galactic dust layer typically have been derived from photometry and other data of stars embedded into or 
seen through this layer.
However, inaccurate distances and photometry cause these properties to still be scarcely understood.

The best data available before the {\it Gaia} mission \citep{gaiabrown, gaiaevans}, 
such as the {\it Hipparcos} parallaxes \citep{hip2}, cover only a hundred parsecs from the Sun.
Therefore, they provide no precise data for determining (i) the dust layer scale height, 
(ii) related extinction through the whole dust half-layer below or above the Sun and, consequently,
(iii) extinction to high-latitude objects situated behind the layer.

As a result, various estimates of these quantities appear to be contradictory.
For example, \citet{gm2018} recently revised the reddening across the whole dust layer. 
We used data from the {\it Gaia} Data Release 1 (DR1) Tycho--Gaia Astrometric Solution (TGAS; \citealt{tgas}) for giants within 415 pc from the Sun, 
in combination with their extinction/reddening estimates from various maps and models
\footnote{A map is a representation in a tabular view, while a model is a representation given by some formulas.}. 
We compared the position of these giants in the Hertzsprung--Russell (HR) diagrams among the isochrones from PAdova and TRieste 
Stellar Evolution Code (PARSEC; \citealt{marigo2017}\footnote{\url{http://stev.oapd.inaf.it/cgi-bin/cmd}}) and
MESA Isochrones and Stellar Tracks (MIST; \citealt{paxton2011, paxton2013, choi2016, mist}
\footnote{\url{http://waps.cfa.harvard.edu/MIST/}}).
The TRILEGAL  Galaxy model \citep{trilegal} with its parameters widely varied was also used.
We found that, given a reliable solar metallicity, the median reddening at Galactic latitudes $|b|>50\degr$, far from the Galactic mid-plane, is $E(B-V)>0.04$ mag, 
with its most probable estimate $E(B-V)=0.06$ mag.
This evidence agrees with the estimates of \citet{teerikorpi1990}, \citet[][hereafter G17]{g17}, and \citet{polarization}.
However, this contrasts with the estimates within $0.01<E(B-V)<0.02$ mag from the widely used 2D reddening maps of \citet[][hereafter SFD98]{sfd} and 
\citet[][hereafter MF15]{2015ApJ...798...88M} and 3D reddening maps of \citet[][hereafter DCL03]{drimmel} and \citet[][hereafter LVV19]{lallement2019}.
Thus, the reddening across the whole dust layer needs to be further explored.

Besides that, different authors provide different estimates of the dust layer scale height \citep[][pp. 469--472]{perryman}.
For example, they include: $35-55$ \citep{vergely1998}, $<70$ \citep{juric}, $\approx100$ \citep[][hereafter G12]{av}, $140$ \citep{bmg1}, and $188$~pc \citep{drispe}.

The dust that provides a considerable reddening exists much behind the scale height of the dust layer.
For example, 5 per cent of dust (and corresponding reddening) is behind the three scale heights if the dust vertical distribution is described by an exponential law: 
$\sim\mathrm{e}^{-|Z-Z_0|/Z_\mathrm{A}}$ \citep[][ p. 265]{parenago}, where $Z$ is the Galactic rectangular coordinate directed to the North Galactic pole,
$Z_\mathrm{A}$ is the scale height of the dust layer and $Z_0$ is the vertical offset of the dust layer mid-plane w.r.t. the Sun.
Hence, to take into account more than 95 per cent of the reddening across the whole dust layer, we have to consider stars within, at least, $3Z_\mathrm{A}$, 
i.e. $|Z|<188\times3=564$ pc. Consequently, the distances to them need to be accurately estimated.

Moreover, there may exist a degeneracy between the reddening and the vertical (i.e. along $|Z|$) gradient of a star intrinsic (dereddened) colour, as well as 
between the extinction and the vertical gradient of a star absolute magnitude.
To avoid this degeneracy, we should consider a wide range of $|Z|$ in a space beyond the Galactic dust layer, i.e., at least,
for $500<|Z|<1000$ pc. In this space, the gradients are high, but the variations of the reddening and extinction are minimal.

At such high $|Z|$, {\it Gaia} DR2 is the first all-sky source of accurate parallaxes. It allows us to consider a complete sample of rather luminous stars from 
{\it Gaia} DR2 in such a space. It must be a common type of stars with accurate photometry in the whole space under consideration.
These stars must be easily selected in the HR diagram using their photometry and parallaxes.
Therefore, we consider the red giant clump domain in the HR diagram.

This domain contains a mix of giants of the clump, branch, and asymptotic branch.
They differ by the nuclear fusion inside them: core helium for the clump, envelope hydrogen for the branch, and both constituents for the asymptotic branch.
Among them, the red clump giants are suitable for our study, being `standard candles' with rather small and predictable variations of their intrinsic colours and 
absolute magnitudes \citep{girardi2016}. However, empirical estimates of the clump intrinsic colours and absolute magnitudes contradict each other
\citep[][hereafter RBA18]{ruiz2018}.
Moreover, they show a large discrepancy with their theoretical predictions \citep{girardi2016}. Thus, the clump intrinsic characteristics need further investigation.

\citet{rcg, montecarlo} has shown that a sample of all stars in the clump domain, selected by use of accurate data, contains
a majority of red clump giants and a minority of branch and asymptotic branch giants.
Then a pure sample of red clump giants can be created by removing branch and asymptotic branch contaminants,
if one uses some assumptions and/or additional photometric, spectroscopic or asteroseismic data
(see, e.g. \citealt{montecarlo, chen2017}).

However, in this study we use another approach thanks to high {\it Gaia} DR2 precision.
Instead of cleaning the sample of all stars in the clump domain, we consider modes of their observables.
To obtain the modes, we round the observables up to 0.01 mag and find the tops of their histograms in each spatial cell.
In the case of multimodal histograms, the lowest value is selected, since it is more probable for the unreddened or 
slightly reddened clump.
In this approach, we follow \citet{g2017}. 
He has shown that the modes of colours, dereddened colours, magnitudes and absolute magnitudes for such a sample are completely 
defined by red clump giants.
In contrast, their mean and median vary considerably due to the influence of branch and asymptotic branch giants.

The aim of this paper is to derive with minimal assumptions
(i) reddening and extinction estimates across the whole Galactic dust layer by use of the red clump giants as extinction 
probes,
(ii) dereddened colours and absolute magnitudes of the red giant clump, and
(iii) some vertical gradients of these quantities due to vertical gradients of age and metallicity of the clump.

This paper is organized as follows. 
In Sect.~\ref{data} we select a sample of {\it Gaia} DR2 giants. 
In Sect.~\ref{direct} we analyse the spatial variations of the modes of some observables in order to estimate 
the extinction and reddening along the Galactic mid-plane and across the dust layer, as well as
the intrinsic colours and absolute magnitudes of the red clump giants.
We compare them with both empirical and theoretical estimates and derive an age--metallicity relation (AMR) in Sect.~\ref{theor}.
We discuss different estimates of the reddening across the whole dust layer in Sect.~\ref{discuss}.
We summarize our findings and state our conclusions in Sect.~\ref{conclusions}.

\section{Data}
\label{data}

To study the properties of the red clump giants in the Galactic dust, we consider a properly constrained space. The space under consideration is limited by:
(i) the precision of the {\it Gaia} DR2 parallaxes, which should be better than 10 per cent,
(ii) the predictability of the properties of the dust layer and the stars embedded within it, despite their variations along the $X$, $Y$ and $Z$ Galactic coordinates 
(see, e.g. \citealt{drispe}),
(iii) the need for a wide range of $Z$.
Therefore, we consider a space cylinder with a radius of 700~pc around the Sun, elongated up to $|Z|<1800$~pc along the $Z$-axis.

This space is especially appropriate to avoid the degeneracy between the reddening and intrinsic gradients.
The reddening is determined more accurately in regions far from the Sun, near the Galactic mid-plane, while the 
gradients are determined more accurately in regions above or below the Sun, far from the mid-plane.

\begin{figure*}
\includegraphics{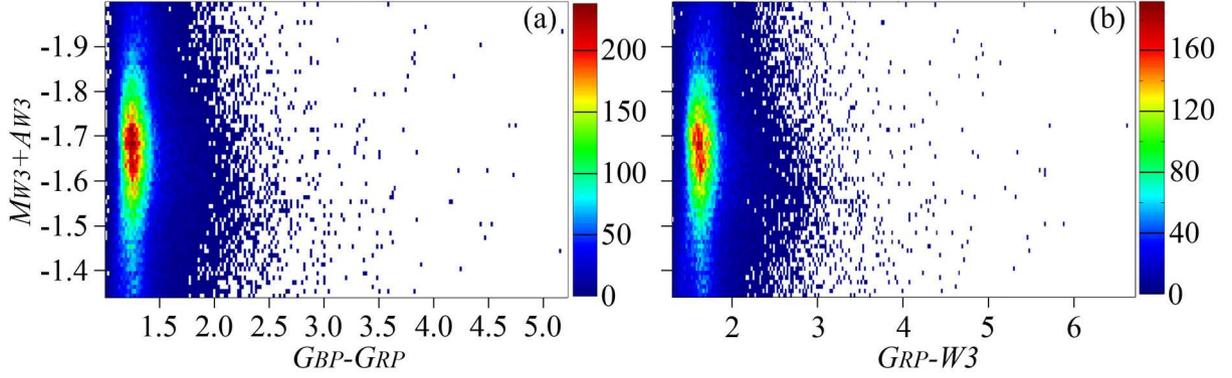}
\caption{The initial HR diagrams (a) `$G_\mathrm{BP}-G_\mathrm{RP}$ versus $M_\mathrm{W3}+A_\mathrm{W3}$' and 
(b) `$G_\mathrm{RP}-W3$ versus $M_\mathrm{W3}+A_\mathrm{W3}$' for the selected sample.
The bins are 0.02 and 0.01 mag for the abscissas and ordinates, respectively. The number of stars in each bin is shown by the colour scale on the right.
}
\label{hr}
\end{figure*}

To derive robust properties of the dust layer, we need a rather long wavelength baseline, i.e. with optical and infrared (IR) bands.
\citet[table 1]{g2016} has estimated the distance ranges where a sample of red clump giants with an accurate photometry in 
various bands is complete.
In the space cylinder under consideration, the {\it Wide-field Infrared Survey Explorer (WISE}, allWISE, \citealt{wise}) $W3$ 
photometry at 10.8 microns provides the most complete sample of the red giant clump among all bands of IR all-sky surveys.
Its photometric precision is $<0.06$ mag for all stars, with a median precision of $0.017$ mag.
By use of the Two Micron All-Sky Survey (2MASS, \citealt{2mass}) $J$, $H$, $K_s$ and {\it WISE} $W1$, $W2$, $W4$ 
photometry of the same precision, we would lose about 2, 10, 2, 43, 8 and 82 per cent of the stars with the $W3$ 
photometry, respectively.
Moreover, the saturation of all nearby giants in the $J$, $H$, $K_s$, $W1$ and $W2$ bands would lead to undesirable removal of 
giants with a little or zero reddening.

Finally, we use the $G_\mathrm{BP}$ and $G_\mathrm{RP}$ bands from {\it Gaia} DR2 in combination with the $W3$ band from {\it WISE}.
We use the {\it WISE} versus {\it Gaia} DR2 cross-identification in order to select all stars in the clump domain of the HR diagram.

The domain must be constrained in order to keep the main concentration of the giants, i.e. their clump, well inside this domain,
whether the extinction and reddening are taken into account or not.
In such a case, no correction for reddening and extinction has to be applied during the sample selection.
To fulfill this, any variation of modes of the observables due to reasonable variations of age, metallicity, 
reddening and extinction law within the space under consideration should keep them within the selection domain.
To constrain the domain, we use reliable data sources, such as TRILEGAL, 3D reddening map of \citetalias{drimmel}
\footnote{The \citetalias{drimmel} estimates are calculated by use of the code of \citet{bovy2016}, 
\url{https://github.com/jobovy/mwdust}}, and extinction law of \citet[][hereafter WC19]{wang2019}
\footnote{Hereafter, using the law of \citetalias{wang2019}, we take into account the dependencies of the extinction 
coefficients on the spectral energy distribution and the extinction itself by use of the curves from figure 2 of \citetalias{wang2019}.
However, as seen from that figure, for the vast majority of our stars this changes $A_\mathrm{G_\mathrm{BP}}$ and 
$A_\mathrm{G_\mathrm{RP}}$ by only a few per cent, while $A_\mathrm{W3}$ does not change at all.
Therefore, our results and conclusions are not affected by these dependencies.}.
The domain is constrained as:
\begin{equation}
\label{select1}
G_\mathrm{BP}-G_\mathrm{RP}>1\,,
\end{equation}
\begin{equation}
\label{select2}
G_\mathrm{RP}-W3>1.3\,,
\end{equation}
\begin{equation}
\label{select3}
-2<W3+5+5\,\log_{10}\varpi<-1.35\,,
\end{equation}
where $\varpi$ is the parallax from {\it Gaia} DR2.
Hereafter, we also use the distances $R$ derived from $\varpi$ by \citet{bailer2018}.
Equation (\ref{select3}) is equivalent to 
\begin{equation}
\label{select4}
-2<M_\mathrm{W3}+A_\mathrm{W3}<-1.35\,,
\end{equation}
taking into account 
\begin{equation}
\label{equw3}
W3+5+5\,\log_{10}\varpi=M_\mathrm{W3}+A_\mathrm{W3}\,,
\end{equation}
where $M_\mathrm{W3}$ and $A_\mathrm{W3}$ are absolute magnitude and extinction in $W3$, respectively. 
Similarly to observable (\ref{equw3}), we can consider the observables
\begin{equation}
\label{equbp}
G_\mathrm{BP}+5+5\,\log_{10}\varpi=M_\mathrm{G_\mathrm{BP}}+A_\mathrm{G_\mathrm{BP}}\,,
\end{equation}
\begin{equation}
\label{equrp}
G_\mathrm{RP}+5+5\,\log_{10}\varpi=M_\mathrm{G_\mathrm{RP}}+A_\mathrm{G_\mathrm{RP}}\,.
\end{equation}

We remove stars with a negative parallax or with the fractional parallax uncertainty (the ratio of the parallax uncertainty to 
the parallax) higher than 0.354, since they have no precise distance \citep{bailer2015}.

Since the modes of observables (\ref{select1}), (\ref{select2}), (\ref{equw3}), (\ref{equbp}) and (\ref{equrp})
are always inside the selection domain, the following relations are true for the red giant clump:
\begin{equation}
\label{equmbp}
mode(M_\mathrm{G_\mathrm{BP}}+A_\mathrm{G_\mathrm{BP}})=mode(M_\mathrm{G_\mathrm{BP}})+A_\mathrm{G_\mathrm{BP}}\,,
\end{equation}
\begin{equation}
\label{equmrp}
mode(M_\mathrm{G_\mathrm{RP}}+A_\mathrm{G_\mathrm{RP}})=mode(M_\mathrm{G_\mathrm{RP}})+A_\mathrm{G_\mathrm{RP}}\,,
\end{equation}
\begin{equation}
\label{equmw3}
mode(M_\mathrm{W3}+A_\mathrm{W3})=mode(M_\mathrm{W3})+A_\mathrm{W3}\,,
\end{equation}
\begin{equation}
\label{equbprp}
mode(G_\mathrm{BP}-G_\mathrm{RP})=mode((G_\mathrm{BP}-G_\mathrm{RP})_0)+E(G_\mathrm{BP}-G_\mathrm{RP})\,,
\end{equation}
\begin{equation}
\label{equrpw3}
mode(G_\mathrm{RP}-W3)=mode((G_\mathrm{RP}-W3)_0)+E(G_\mathrm{RP}-W3)\,,
\end{equation}

We remove 98 stars with the {\it Gaia} DR2 
\verb"phot_bp_rp_excess_factor"$\equiv (F_\mathrm{BP}+F_\mathrm{RP})/F_\mathrm{G}>1.3+0.06(G_\mathrm{BP}-G_\mathrm{RP})^2$,
where $F_\mathrm{BP}$, $F_\mathrm{RP}$, and $F_\mathrm{G}$ are the fluxes within the {\it Gaia} filters.
The reason of their removal is that they are not `well-behaved single sources' \citep{gaiaevans}.

172 stars appear outliers in the diagram $G_\mathrm{BP}-G_\mathrm{RP}$ versus $G_\mathrm{RP}-W3$, apparently, 
due to a wrong {\it Gaia} DR2 -- {\it WISE} cross-identification.
We remove these outliers by use of our empirical relation for the remaining stars: $G_\mathrm{RP}-W3<0.86+1.3(G_\mathrm{BP}-G_\mathrm{RP})$.

The spatial distribution of the rejected stars is quite uniform, therefore their removal does not influence our results.

\begin{table*}
\def\baselinestretch{1}\normalsize\normalsize
\caption[]{The main data for the selected sample:
source\_id, right ascension and declination for the epoch 2015.0 (deg), parallax (mas), standard error of parallax (mas), 
Galactic longitude and latitude for the epoch 2015.0 (deg), distance $r\_est$ from \citet{bailer2018} (parsec), $G_\mathrm{BP}$ (mag), $G_\mathrm{RP}$ (mag),
and $W3$ from the allWISE catalogue (mag). All the data are from {\it Gaia} DR2, except $r\_est$ and $W3$. The complete table is available online.
}
\label{master}
\[
\begin{tabular}{rrrrrrrrrrr}
\hline
\noalign{\smallskip}
Source\_id          & RA   & DEC & $\varpi$ & e\_$\varpi$ & $l$ & $b$ & $r\_est$ & $G_\mathrm{BP}$ & $G_\mathrm{RP}$ & $W3$ \\
\hline
\noalign{\smallskip}
1000049810224526080 & 104.450301 & 55.980873 & 2.83 & 0.04 & 160.365533 & 23.064404 & 350.18 & 9.453  & 8.071 & 6.290 \\
1000134678777721088 & 103.017215 & 55.776143 & 3.60 & 0.05 & 160.348636 & 22.235018 & 275.68 & 8.668  & 7.323 & 5.570 \\
1000344651138651776 & 102.573738 & 56.874385 & 2.85 & 0.11 & 159.134517 & 22.299573 & 347.67 & 8.686  & 7.439 & 5.817 \\
1000594171558726656 & 104.647456 & 57.390965 & 1.28 & 0.05 & 158.907905 & 23.520080 & 762.58 & 10.347 & 9.293 & 7.956 \\
1000598668388532864 & 104.438721 & 57.496416 & 1.28 & 0.04 & 158.766577 & 23.436812 & 762.24 & 10.615 & 9.314 & 7.615 \\
\ldots              & \ldots     & \ldots    & \ldots & \ldots & \ldots & \ldots    & \ldots & \ldots & \ldots & \ldots \\
\hline
\end{tabular}
\]
\end{table*}


The final sample contains 101\,810 giants.
In Table~\ref{master} we list some important information on the selected sample.
The median relative error of $R$ in the sample is 2 per cent. 
Only 720 stars (0.7 per cent) have a relative error of $R$ larger than 10 per cent.
The photometry of the selected stars is very precise: 
$\sigma(G_\mathrm{BP})<0.05$ and $\sigma(G_\mathrm{RP})<0.03$ mag for all the stars,
while the medians are $\sigma(G_\mathrm{BP})=\sigma(G_\mathrm{RP})=0.001$ mag.

The distribution of the sample inside the selection domain in the HR diagrams is shown in Fig.~\ref{hr}.

\begin{figure}
\includegraphics{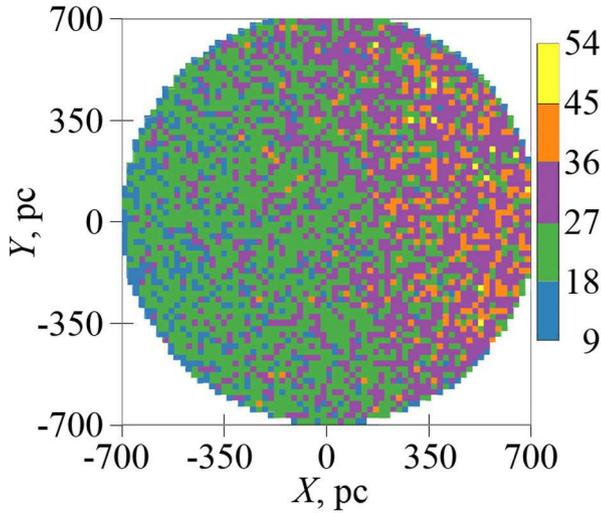}
\caption{The sample projected into the $XY$ plane in the bins of $20 \times 20$~pc. The number of stars in each bin is shown 
by the colour scale on the right. The Sun is at the centre. The Galactic Centre is to the right.
}
\label{xy}
\end{figure}

\begin{figure}
\includegraphics{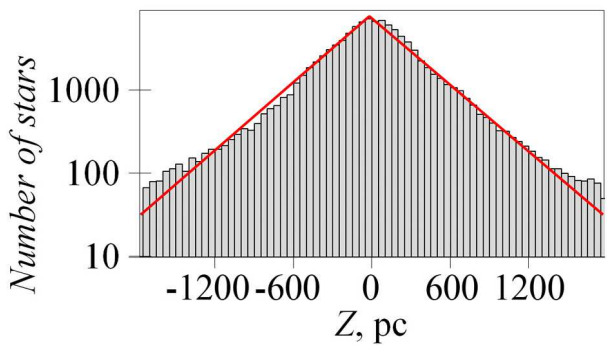}
\caption{The distribution of the sample by $Z$.
The most reliable exponential approximation of this distribution with a scale height of 325 pc is shown by the red line.
}
\label{distz}
\end{figure}

The completeness of the sample is evident from Fig.~\ref{xy} where the sample is projected into the $XY$ plane 
in the bins of $20 \times 20$ pc.
It shows no artifact and no decrease to the periphery, but only an expected gradient to the Galactic Centre (to the right).

The distribution of the sample by $Z$ is shown in Fig.~\ref{distz}.
This can be compared with figure 6 of \citet{rcg}. The most reliable exponential approximation of this distribution,
with a scale height of 325~pc and a maximum spatial density at 20~pc south from the Sun, is shown in Fig.~\ref{distz} by the red line.
Accordingly, both the mean and median values of $Z$ in the sample is $-22$~pc.
This deviation from zero reflects the shift of the Galactic mid-plane from the Sun.
The scale height and the offset of the Sun differ from the previous estimates obtained by \citet{rcg}, 280 and 13 pc,
respectively, due to a higher completeness of the current sample.
The median $|Z|$ of the sample is 212~pc.
73 and 90 per cent of the sample are within $|Z|<400$ and $<700$~pc, respectively.
Consequently, the majority of stars seem to be inside the dust layer.

To consider the spatial distribution of the sample in detail, one has to separate the clump, branch and asymptotic branch
giants from their mix. 
However, this is out of the scope of this paper.

\section{Spatial variations of the observables}
\label{direct}

\begin{figure}
\includegraphics{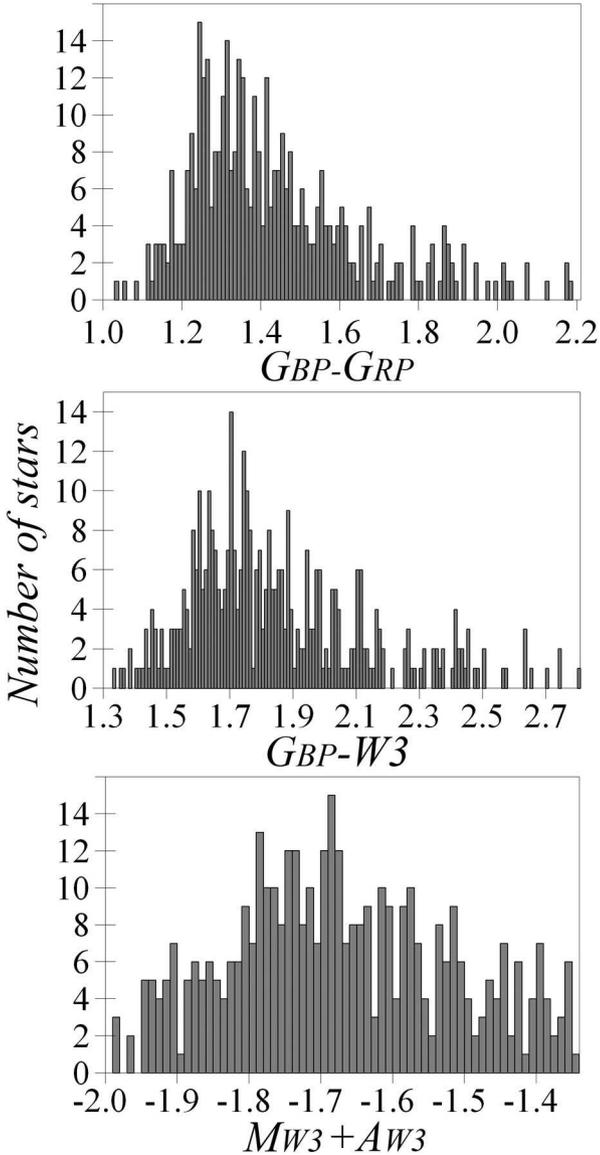}
\caption{Histograms of some observables in a spatial cell with 400 giants from our sample.
}
\label{cell}
\end{figure}

The main assumptions of our approach are defined by equations (\ref{equmbp})--(\ref{equrpw3}).
Apart from those, we use the following:
\begin{enumerate}
\item Any spatial variations of the dereddened colours and absolute magnitudes of the clump can be presented as their linear 
vertical and radial gradients.
The former is a gradient along $|Z|$, while the latter is a gradient along the galactocentric position.
This position, w.r.t. the solar circle, is calculated as $X'=R_{\sun}-((R_{\sun}-X)^2+Y^2+Z^2)^{0.5}$, 
where $R_{\sun}=8200$ pc is the adopted distance from the Sun to the Galactic Centre \citep[][pp. 495--496]{perryman}.
We define $X'$ to be positive towards the Galactic Centre. Our space cylinder covers the range $-868<X'<699$ pc.
The uncertainty of $R_{\sun}$ has a negligible effect on our results, since we consider a Galactic vicinity of the Sun.
For the same reason, $X'$ is close to $X$.
\item The dust volume density has a very shallow gradient with $|Z|$ in regions far from the Galactic mid-plane.
More precisely, cumulative extinction and reddening have a negligible increase with $|Z|$ at $|Z|>400$ pc.
\end{enumerate}

To analyse the spatial variations of observables (\ref{equmbp})--(\ref{equrpw3}), we need to calculate their modes in some small 
spatial cells.
As mentioned in Sect.~\ref{intro}, to obtain the modes, we round the observables up to 0.01 mag and find the tops of their 
histograms in each cell.
An example of the histograms for a cell with a rather high extinction and reddening is shown in Fig.~\ref{cell}.
It shows long red tails of $G_\mathrm{BP}-G_\mathrm{RP}$ and $G_\mathrm{RP}-W3$ due to reddening and,
what is less important, due to the branch and asymptotic branch contaminants.
The observed asymmetry of the $M_\mathrm{W3}+A_\mathrm{W3}$ histogram is caused by the contaminants.
Despite this, the modes are determined fairly confidently.

\begin{figure*}
\includegraphics{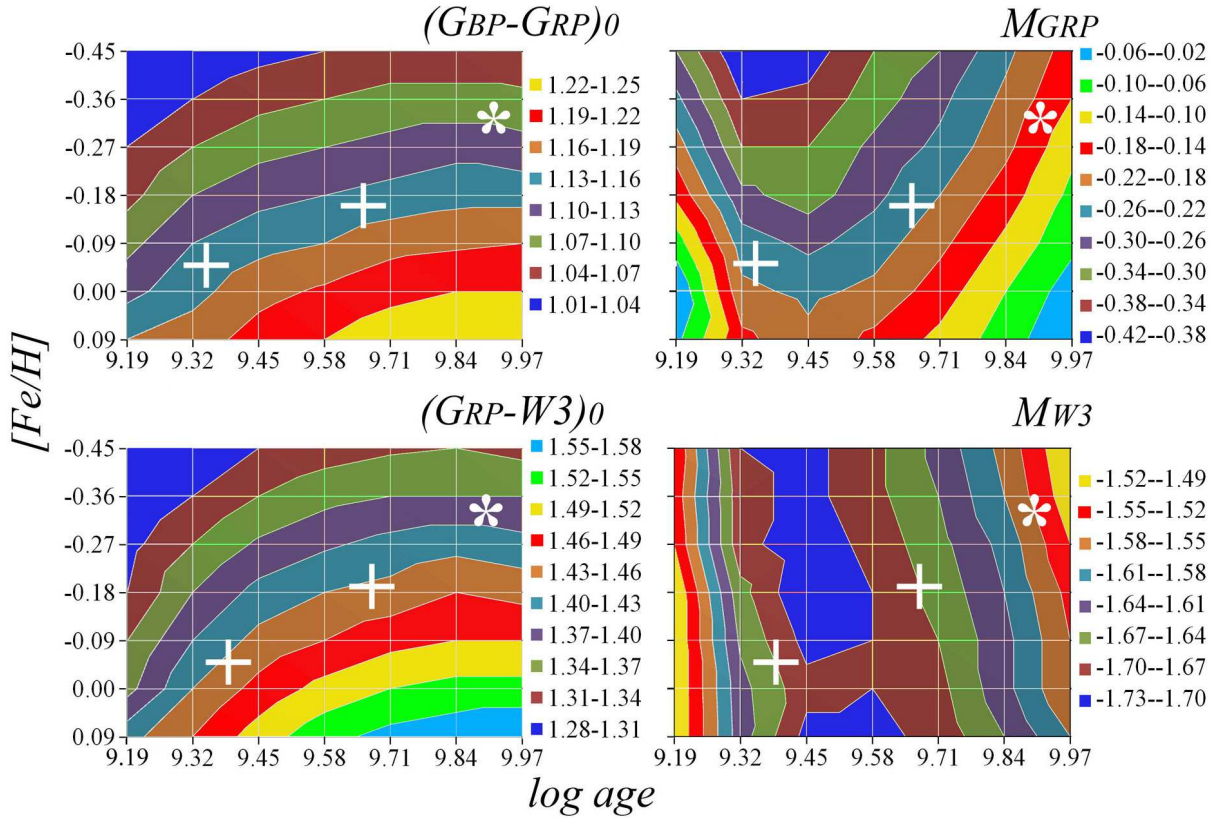}
\caption{PARSEC predictions of the intrinsic colours and absolute magnitudes of the clump as functions of age and [Fe/H].
The solutions from Table~\ref{hrrelation} for the input from Table~\ref{zero} are shown by the white crosses and stars
for $Z=0$ and $|Z|=1700$ pc, respectively.
}
\label{parsec}
\end{figure*}

\begin{figure*}
\includegraphics{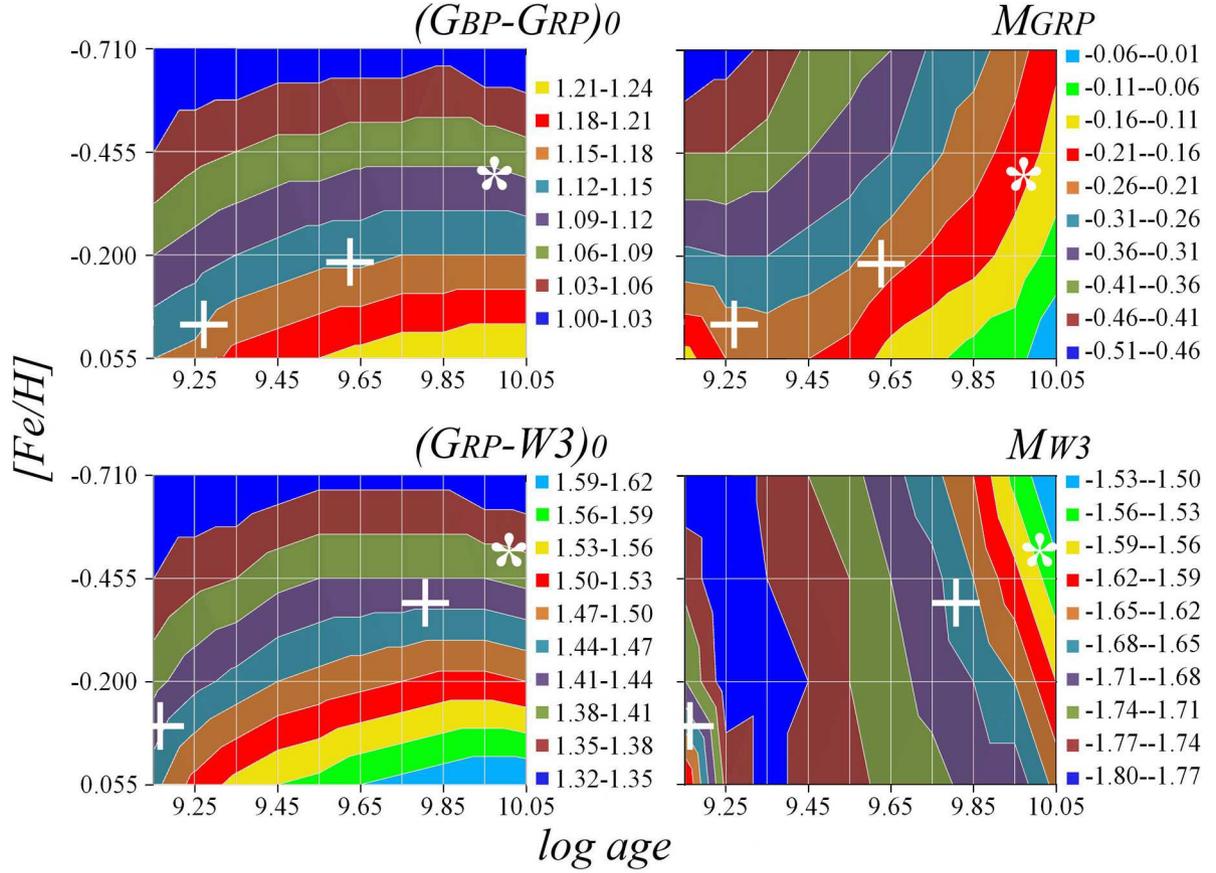}
\caption{The same as Fig.~\ref{parsec} but for MIST.
}
\label{mist}
\end{figure*}

\begin{figure*}
\includegraphics{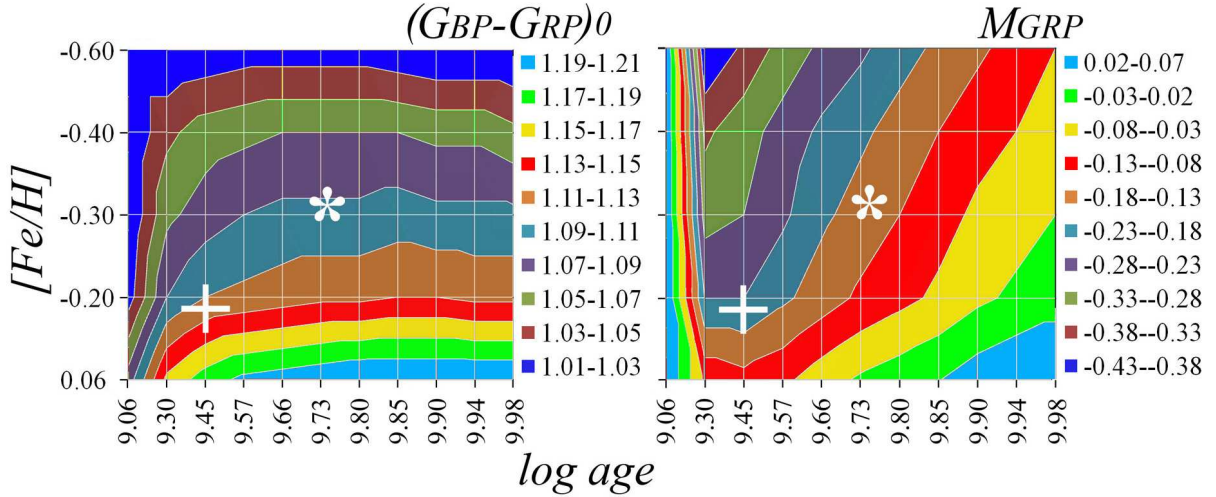}
\caption{The same as Fig.~\ref{parsec} but for IAC-BaSTI.
}
\label{iacbasti}
\end{figure*}

Various uncertainties contribute to the total uncertainty of the modes.     Three uncertainties seem to be most important:
(i) the photometric uncertainty,
(ii) the uncertainty due to random fluctuations of the stellar content in the cell with higher or lower percentage of the
clump giants w.r.t. the contaminants, and
(iii) the uncertainty due to natural fluctuations of the interstellar medium between us and the stars in the cell, 
followed by related fluctuations of the extinction, reddening and observables.

We expect that the uncertainties (i) and (ii) decrease with an increasing number of stars in a cell and, in turn, with an 
increasing size of the cell.
In contrast, the uncertainty (iii) has a little, if any, decrease with an increasing size of the cell,
since the medium fluctuations increase with the size (see, e.g. \citealt{g2019}).
Hence, an optimal number of stars in a cell (and an optimal cell size) must exist.
With this or greater number of stars in a cell, the uncertainty (iii) dominates in the total uncertainty.
We select the optimal number of stars in a cell empirically: by increasing it until the standard deviation of the modes 
of the observables in adjacent spatial cells is stabilized.
The optimal number appears to be 400 stars.

We can consider the standard deviation of the modes, since the fluctuations of the medium from cell to cell are random.
Consequently, with the dominance of the uncertainty (iii), we expect to see a random scatter of the modes of the observables
in several independent cells.

The formal precision of the derived modes is 0.01 mag. However, their accuracy is defined by the medium fluctuations.
This accuracy varies from 0.01 mag for cells near the Sun and at high-latitudes up to several hundredths of a 
magnitude in distant cells near the Galactic mid-plane \citep{g2019}.
Consequently, we expect the uncertainty of an average or a gradient, derived from many cells, 
to be at a level of $0.01$ mag.

A sample of unreddened giants is needed to derive their intrinsic characteristics and also as a reference for reddened giants.
However, we cannot create such a sample from the nearest giants. Indeed, to select 400 nearest giants,
we have to consider a vicinity of 111 pc around the Sun.
However, \citet{polarization} show that the extinction and reddening are the lowest, yet, not negligible at an accuracy level 
of 0.01 mag even in the Local Cavity, or Local Bubble, a region within about 80~pc from the Sun.
Therefore, any extinction or reddening estimate in this space should be considered only as an upper limit of the estimate at $R=0$.
Moreover, due to the nearly uniform spatial distribution of the sample giants near the Galactic mid-plane, half of them, being 
within $R<111$ pc have, in fact, $92<R<111$ pc.  Naturally, they have a typical reddening of several hundredths of a magnitude.
Considering such slightly reddened stars as unreddened ones, one would introduce a bias to any further estimate of the reddening 
for distant stars.
This bias is especially important at high-latitudes where it is comparable to the observed low reddening and, therefore, 
makes it very uncertain.

\subsection{The thin coordinate layer}
\label{thinlayer}

We can estimate the dereddened colours and absolute magnitudes for nearby clump giants by extrapolating the moving modes of the 
observables to $R=0$.
The moving modes are calculated using a window of 400 stars. We consider a thin coordinate layer of $|Z|<40$~pc with 10\,802 stars.

The thickness of this coordinate layer is set in a way that it certainly contains a significant mass of dust.
Also the coordinate layer must contain the mid-plane of the dust layer, despite the shift $Z_0$ of this mid-plane from the Sun.
This shift is known worse than the shifts of the Sun from some stellar distribution mid-planes and from the
geometrical Galactic mid-plane. However, this shift can be estimated as $10<Z_0<30$ pc (see the discussion in \citealt{ob}).
Therefore, the thickness of the thin coordinate layer as $|Z|<40$~pc seems to be appropriate.
Similarly, \citet{vergely1998} have considered the reddening within a thin coordinate layer of $|Z|<40$~pc 
in order to derive some properties of the dust medium near the Galactic mid-plane.

In total, we calculate $10\,802-400=10\,402$ moving modes. Only $10\,802/400\approx27$ of them are independent from each other.
We apply a resampling method to check the robustness of our results\footnote{Hereafter we apply such a verification of our 
results using a resampling method in all cases when we calculate moving modes.}.
Namely, we reproduce our results with many randomly selected subsamples of 27 modes out of 10\,402.
The results for these realizations of the resampling method appear the same as those for the 10\,402 moving modes.

The final precision of the extrapolation at a level of $<0.015$ mag is ultimately defined by the standard deviation of the modes 
at a level of $<0.08$ mag due to the medium fluctuations and by the number (27) of the independent modes: $0.08/27^{0.5}\approx0.015$.

We should obtain the radial gradients of the dereddened colours and absolute magnitudes using radial gradients of age and metallicity.
In the literature, no considerable radial age gradient is known in the thin layer under consideration.
The estimate of the radial metallicity\footnote{[Fe/H] is the logarithmic Fe abundance relative to the Sun.} gradient  
\begin{equation}
\label{radialgradient}
\Delta \mathrm{[Fe/H]}/\Delta X'=-0.082$ dex kpc$^{-1}
\end{equation}
for the clump giants within $|Z|<100$ and $-3500<X'<2000$ pc from \citet{huang2015} seem to be the most appropriate for our 
subsample in the thin layer.
Other estimates for such a gradient are within a few hundredths of dex kpc$^{-1}$ from (\ref{radialgradient})
(see, e.g. \citealt{onaltas2016}).
A reasonable estimate of the median metallicity of clump giants at $X'=0$ is $-0.10<\mathrm{[Fe/H]}<0$ \citep{huang2015}.
In combination with gradient (\ref{radialgradient}), this suggests that a reasonable estimate of the median metallicity in our 
thin layer is within the range
\begin{equation}
\label{medianmetal}
-0.16<\mathrm{[Fe/H]}<+0.06\,.
\end{equation}

To convert a gradient of metallicity into gradients of the dereddened colours and absolute magnitudes, we should use some 
theoretical stellar evolution model estimates.
They are based on theoretical isochrones in combination with colour versus effective temperature ($T_\mathrm{eff}$) relations and 
bolometric corrections.
We use the model estimates from PARSEC\footnote{PARSEC version 1.2S with $\mathrm{[\alpha/Fe]}=0$, solar metallicity $Z=0.0152$, 
mass-loss efficiency $\eta=0.2$, where $\eta$ is the free parameter in the Reimers' law \citep{reimers}.},
MIST\footnote{MIST version 1.2 with $\mathrm{[\alpha/Fe]}=0$ and the reference protosolar metallicity $Z=0.0142$.}, 
and the Bag of Stellar Tracks and Isochrones (IAC-BaSTI, \citealt{newbasti}
\footnote{\url{http://basti-iac.oa-abruzzo.inaf.it/index.html}.
This IAC-BaSTI version assumes $\mathrm{[\alpha/Fe]}=0$, overshooting, diffusion, mass-loss efficiency $\eta=0.3$.
IAC-BaSTI does not provide any estimate for the $W3$ band.}).
The predictions of the intrinsic colours and absolute magnitudes as functions of age and [Fe/H] are presented in Fig.~\ref{parsec}, 
\ref{mist} and \ref{iacbasti} for PARSEC, MIST and IAC-BaSTI, respectively.

These predictions will be compared with our empirical estimates in Sect.~\ref{iso}.

We note that the different models provide fairly comparable predictions for $(G_\mathrm{BP}-G_\mathrm{RP})_0$ and 
$M_\mathrm{G_\mathrm{RP}}$.
In contrast, the PARSEC predictions for $(G_\mathrm{RP}-W3)_0$ and $M_\mathrm{W3}$ are, respectively, systematically bluer and 
fainter  by several hundredths of a magnitude than those from MIST.
This is especially noticeable in the most interesting region $2.5-9.3$ Gyr and $-0.45<\mathrm{[Fe/H]}<0.05$.
We will discuss this in more detail in Sect.~\ref{iso}.

To estimate the gradients, we should consider a proper range of ages.
\citet{girardi2016} notes that `in any galaxy with a relatively constant star-formation rate over gigayear scales, 
the mean age of the red clump will be relatively young and located, indicatively, somewhere between 1 and $\sim4$ Gyr'.
The upper limit is still rather uncertain. Therefore, we conservatively adopt it as 6 Gyr.
Considering the lower limit, we must take into account the division of the clump giants into two main types \citep{girardi2016}.
Low-mass clump gaints have degenerate helium cores at the RGB stage, while the cores of high-mass clump giants are never
completely degenerated.
As a result, low-mass clump giants have similar masses of their helium cores at the stage of the helium nuclear fusion and, hence,
similar colours and absolute magnitudes. This allows them to form a compact clump in HR diagrams.
In contrast, high-mass clump giants have scattered helium core masses and, hence, scattered absolute magnitudes in HR diagrams.
With slightly bluer colours w.r.t the low-mass clump giants, the high-mass clump giants form an additional `vertical structure' at
the blue side of the main clump in a HR diagram, as discussed by \citet{girardi2016}.

PARSEC, MIST and IAC-BaSTI consistently show that the high- and low-mass clump giants are separated by a mass value of 1.7 solar 
masses.
Due to the correlation between mass and age for clump giants, this means a dominance of high-mass clump giants for an age 
$\lesssim1.6$ Gyr ($\log age\lesssim9.2$), their absence for an age $\gtrsim2$ Gyr ($\log age\gtrsim9.3$) and a mix of clump giants 
for an in-between age.
Fig.~\ref{parsec}, \ref{mist} and \ref{iacbasti} confirm this: the colours and absolute magnitudes of the clump change significantly
for $\log age<9.3$ and, especially, for $\log age<9.2$.

There are observational evidences (e.g. those summarized in \citealt{girardi2016}) that the low-mass clump giants certainly 
dominate in the clump domain of any HR diagram for the solar neighbourhood\footnote{This may not be the case in some other regions 
of the Galaxy or in other galaxies. That is why \citet{girardi2016} sets the lower limit of the mean age of the clump at 1 Gyr.}.
Consequently, similar to the above-mentioned contamination by branch and asymptotic branch giants, a contamination by high-mass 
clump giants cannot affect the modes of the observables. These modes are completely defined by low-mass clump giants.
Therefore, the median and mean age of the clump in the space under consideration must be within $1.6-6$ Gyr ($9.20<\log age<9.78$).
Hence, we consider the age range $<1.6$ Gyr in Fig.~\ref{parsec}, \ref{mist} and \ref{iacbasti} only for illustrative purposes.

Fig.~\ref{parsec}, \ref{mist} and \ref{iacbasti} show that for gradient (\ref{radialgradient}) within range (\ref{medianmetal}) 
and a specific age within $1.6-6$ Gyr these models provide rather consistent gradients of the dereddened colours and absolute 
magnitudes.
Moreover, these gradients are rather small: only a few hundredths of a magnitude.
For example, for 2.8 Gyr ($\log age=9.45$), PARSEC gives $(G_\mathrm{BP}-G_\mathrm{RP})_0=1.15$ and 1.18 
for $\mathrm{[Fe/H]}=-0.09$ and $-0.01$, respectively, and, finally, the colour gradient of 0.03 mag.
For the same input, MIST gives $(G_\mathrm{BP}-G_\mathrm{RP})_0=1.17$ and 1.19 with a gradient of 0.02 mag, while
IAC-BaSTI gives $(G_\mathrm{BP}-G_\mathrm{RP})_0=1.15$ and 1.17 with a gradient of 0.02 mag.
Note that the radial gradients far from the Galactic mid-plane are even lower \citep{huang2015}.

It is worth noting that the gradients of the dereddened colours and absolute magnitudes cannot be derived empirically by use of 
our data.
The reason for that is that both dust volume density and metallicity tend to increase towards the Galactic Centre.
Hence, the extinction, the reddening, as well as the dereddened colours and absolute magnitudes under consideration 
also tend to increase towards the Galactic Centre.
The degeneracy of these quantities makes the uncertainty of any empirical estimate of the gradients unacceptable.

Therefore, we cannot estimate a differential extinction and reddening in the thin layer along $X$ or $X'$,
but only along $Y$, where we find no gradient.

However, in this paper we do not seek to study the differential reddening, but extrapolation of the observables
to $R=0$. To do this, we consider the observables as functions of $R$.
In such a way, the gradients at negative and positive $X'$ compensate each other.
This results in a negligible contribution of the small theoretical gradients into our extrapolation, 
at a level $<0.01$ mag within our thin layer.
Therefore, we can ignore any radial gradient for the thin layer in this extrapolation.
Consequently, any variation of an observable with $R$ is considered as a result of reddening or extinction.
Note that the extrapolation is performed at the expense of the averaging of the longitudinal variations of the observables.

\begin{figure}
\includegraphics{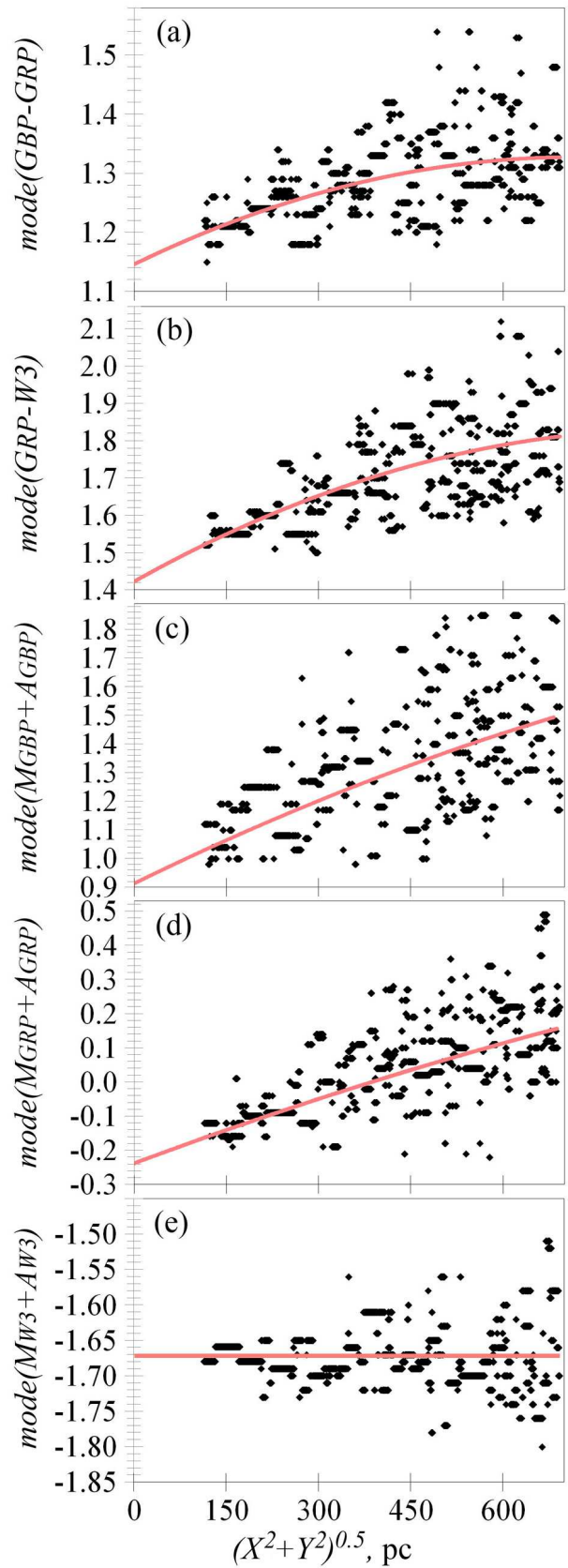}
\caption{Moving modes of the observables, calculated with a window of 400 stars, as functions of $(X^2+Y^2)^{0.5}$ for 
10\,802 stars with $|Z|<40$~pc. The curves show approximations and extrapolations by parabolas.
}
\label{layer40pc}
\end{figure}

\begin{figure}
\includegraphics{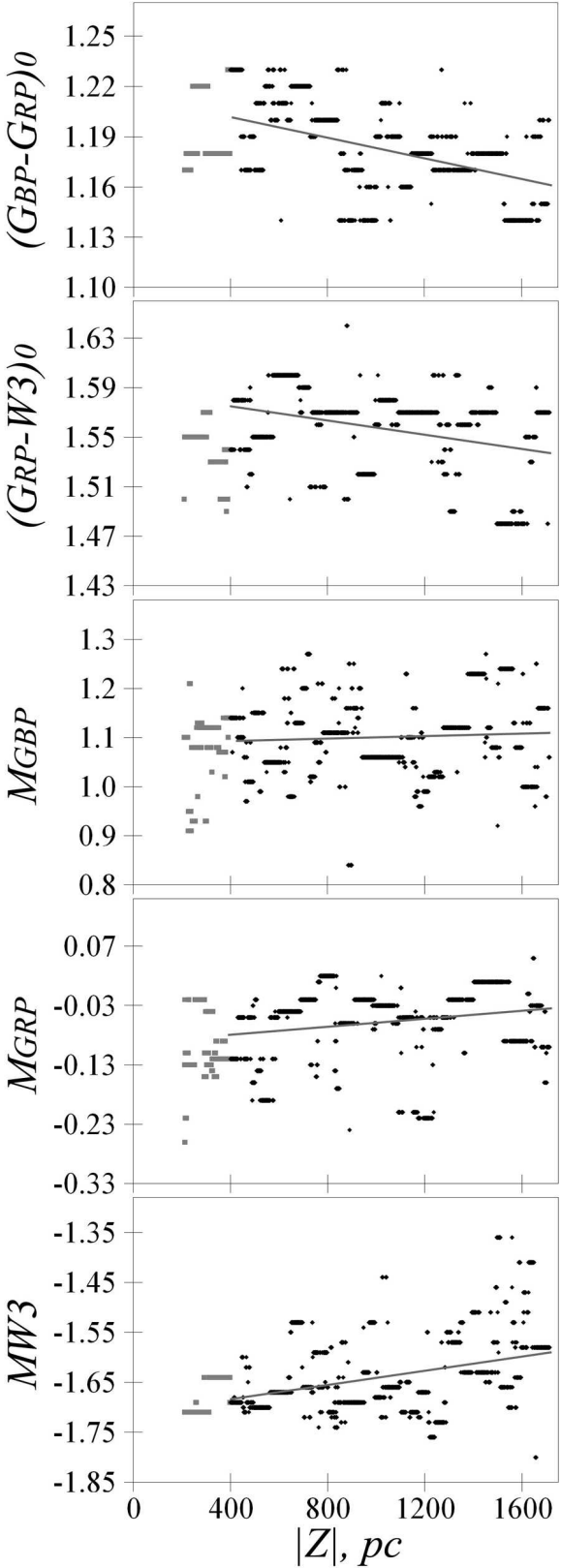}
\caption{The modes of the dereddened colours and absolute magnitudes obtained by use of the reddening from 
\citetalias{2015ApJ...798...88M} and extinction law from \citetalias{wang2019} for 5821 stars within $|b|>68.75\degr$.
The modes for $|Z|$<400 and $|Z|>400$ pc are shown by grey and black symbols, respectively.
The gradients for $|Z|>400$ pc are shown by lines.
}
\label{verticalgradients}
\end{figure}

The moving modes of the observables as functions of $(X^2+Y^2)^{0.5}$ are shown in Fig.~\ref{layer40pc}.
Since the layer is thin, this figure would not change, if we use $R$ instead of $(X^2+Y^2)^{0.5}$.

Fig.~\ref{layer40pc} shows a distinct increase of all the modes with $(X^2+Y^2)^{0.5}$ due to extinction or reddening, 
except $mode(M_\mathrm{W3}+A_\mathrm{W3})$. 
The latter shows no trend due to negligible variations of both $M_\mathrm{W3}$ and $A_\mathrm{W3}$.

The variations in Fig.~\ref{layer40pc} can be approximated by some simple functions to derive the 
dereddened colours, absolute magnitudes, extinctions and reddenings by use of equations (\ref{equmbp})--(\ref{equrpw3}).

We consider a cumulative extinction or reddening in a dust layer with an almost uniform dust distribution 
along the $X$- and $Y$-axes, and a nearly exponential distribution along the $Z$-axis.
In this case, a simple simulation allows us to choose the approximation by a parabolic function of $R$ or 
$(X^2+Y^2)^{0.5}$, which saturates when coming out of the layer.
This was first realized by \citet[][hereafter AGG92]{arenou} [their equation~(5) and figure 1].
They approximated the extinction in 199 celestial areas through some parabolas of $R$ in their 3D extinction model.

\begin{table*}
 \centering
\def\baselinestretch{1}\normalsize\normalsize
\caption[]{The modes of the observables in a thin layer of $|Z|<40$~pc approximated by parabolas and extrapolated 
to $R=0$.
The uncertainty of all the mode values is 0.015 mag.
}
\label{thinlayer40pc}
\begin{tabular}[c]{rrrrrr}
\hline
\noalign{\smallskip}
 $(X^2+Y^2)^{0.5}$, pc & $mode(M_\mathrm{G_\mathrm{BP}}+A_\mathrm{G_\mathrm{BP}})$ & $mode(M_\mathrm{G_\mathrm{RP}}+A_\mathrm{G_\mathrm{RP}})$ & $mode(M_\mathrm{W3}+A_\mathrm{W3})$ & $mode(G_\mathrm{BP}-G_\mathrm{RP})$ & $mode(G_\mathrm{RP}-W3)$ \\
\hline
\noalign{\smallskip}
$0$   & $0.91$ & $-0.24$ & $-1.67$ & $1.15$ & $1.43$ \\ 
$700$ & $1.50$ & $0.16$ & $-1.67$ & $1.34$ & $1.83$ \\ 
\hline
\end{tabular}
\end{table*}


The result of our approximation of the modes by parabolas and their extrapolation to $R=0$ is presented in Table~\ref{thinlayer40pc}.
As a by-product, we derive the following average extinctions and reddenings for $(X^2+Y^2)^{0.5}=700$ and $|Z|<40$ pc as the 
differences between the values at $R=0$ and $(X^2+Y^2)^{0.5}=700$ pc:
\begin{equation}
\label{thinagbp}
A_\mathrm{G_\mathrm{BP}}=1.50\pm0.015-0.91\pm0.015=0.59\pm0.02\,,
\end{equation}
\begin{equation}
\label{thinagrp}
A_\mathrm{G_\mathrm{RP}}=0.16\pm0.015-(-0.24\pm0.015)=0.40\pm0.02\,,
\end{equation}
\begin{equation}
\label{thinaw3}
A_\mathrm{W3}=-1.67\pm0.015-(-1.67\pm0.015)=0.00\pm0.02\,,
\end{equation}
\begin{equation}
\label{thinebprp}
E(G_\mathrm{BP}-G_\mathrm{RP})=1.34\pm0.015-1.15\pm0.015=0.19\pm0.02\,,
\end{equation}
\begin{equation}
\label{thinerpw3}
E(G_\mathrm{RP}-W3)=1.83\pm0.015-1.43\pm0.015=0.40\pm0.02\,.
\end{equation}

Our empirical extinction law, written by results (\ref{thinagbp})--(\ref{thinerpw3}), can be compared with some other extinction laws.
Given $A_\mathrm{G_\mathrm{BP}}=0.59$ mag, the extinction laws of \citet[][hereafter DIB14]{davenport2014} for low latitudes
\footnote{\citetalias{davenport2014} provide different extinction laws for low and high-latitudes.
The study of \citetalias{davenport2014} is very important as the only direct comparison of the empirical extinction laws in the 
Galactic disc and halo: see their table 3 and figures 6 and 7.}, \citet[][hereafter SMS16]{schlafly2016}, and \citetalias{wang2019}
provide $A_\mathrm{G_\mathrm{RP}}=0.35$, 0.35, 0.35, $A_\mathrm{W3}=0.06$, 0.03, 0.02,
$E(G_\mathrm{BP}-G_\mathrm{RP})=0.24$, 0.24, 0.24, and $E(G_\mathrm{RP}-W3)=0.31$, 0.32, 0.33 mag, respectively.
The widely used extinction law by \citet[][hereafter CCM89]{ccm89} does not provide $A_\mathrm{W3}$.
However, for $A_\mathrm{G_\mathrm{BP}}=0.59$ mag the \citetalias{ccm89} extinction law with 
$R_\mathrm{V}\equiv A_\mathrm{V}/E(B-V)=3.1$ gives 
$A_\mathrm{G_\mathrm{RP}}=0.37$ and $E(G_\mathrm{BP}-G_\mathrm{RP})=0.22$ mag
in best agreement with our extinction law.

We derive the colour excess ratios (CERs) $E(G_\mathrm{RP}-W3)/E(G_\mathrm{BP}-G_\mathrm{RP})=2.11^{+0.35}_{-0.30}$, 1.24, 1.28, 
and 1.33, from our results and the laws by \citetalias{davenport2014} for low latitudes,
\citetalias{schlafly2016}, and \citetalias{wang2019}, respectively.

It is seen that these extinction laws differ in their predictions for $A_\mathrm{W3}$. Our extinction law deviates from the remaining 
ones, but no more than $3\,\sigma$.
This deviation can be explained by our averaging of the longitudinal variations of the observables.

Result (\ref{thinagbp}) corresponds to $A_\mathrm{V}/R=0.83$, 0.82, 0.82, 0.84 mag kpc$^{-1}$ by use of the extinction laws 
by \citetalias{ccm89} with $R_\mathrm{V}=3.1$, \citetalias{davenport2014} for low latitudes, \citetalias{schlafly2016}, and
\citetalias{wang2019}, respectively.
Similarly, result (\ref{thinagrp}) corresponds to $A_\mathrm{V}/R=0.87$, 0.94, 0.95, 0.97, respectively.
The average is $A_\mathrm{V}/R=0.88\pm0.07$ mag kpc$^{-1}$.
It is worth noting that this estimate is averaged over the longitude.
Hence, this reflects the differential reddening along $Y$ rather than along $X'$, where the gradients of the observables are high.
This may be a reason why this estimate is lower than common estimates along the mid-plane, e.g.
$1.1<A_\mathrm{V}/R<1.7$ mag kpc$^{-1}$ of \citet[][and references therein]{vergely1998}.
The averaging of the longitudinal variations of the observables makes our estimates (\ref{thinagbp})--(\ref{thinerpw3}) less 
robust and important than the extrapolated estimates of the dereddened colours and absolute magnitudes in Table~\ref{thinlayer40pc}.

\subsection{Vertical gradients}
\label{gradient}

We estimate the vertical gradients of the intrinsic colours and absolute magnitudes in a space beyond the dust layer.
There must be no variations of the cumulative reddening in such a space along each line of sight.
However, such a space maintains some small variations of the cumulative reddening from one line of sight to another 
due to medium fluctuations between these lines of sight inside the dust layer.
Moreover, lines of sight far from the Galactic poles cross rather dense regions of the dust layer.
Hence, we must only consider some neighbourhood of the poles.

We consider a subsample of 5821 stars within the vertical cones resting on the base of the space cylinder under consideration, 
i.e. with $|b|>90\degr-\arctan(700/1800)\approx68.75\degr$.
We consider a space with $|Z|>400$ pc as one with a negligible increase of the reddening along $|Z|$.
Indeed, any reasonable vertical distribution of dust suggests such an increase by only $<12$ per cent.

The observables of each star must be corrected for reddening and extinction through the whole dust half-layer.
Therefore, we should adopt some estimates of the reddening or extinction from us to infinity, i.e. taken from a 2D reddening map.
We use one of the most accurate 2D reddening maps of \citet[][hereafter MF15]{2015ApJ...798...88M}.
It is based on the observations of the dust emission in the far-IR by {\it Planck} and a reddening-to-emission calibration.
In Sect.~\ref{discuss} we discuss that this map may have some systematic errors.
However, here we do not need very accurate estimates of the reddening and/or extinction across the dust layer, 
since they are nearly cancel each other out in the colour and magnitude gradients outside the layer.

Note that 3D reddening/extinction maps and models are not useful in studying the vertical gradients, since, usually, they do not 
extend far enough from the Galactic mid-plane. For example, the 3D maps by
\citet[][hereafter LVV19]{lallement2019}\footnote{\url{http://stilism.obspm.fr}} and 
\citet[][hereafter G17]{g17} extend up to only $|Z|<400$ and 600 pc, respectively.
Only the 3D reddening maps by \citetalias{drimmel} and \citet[][hereafter GSF19]{green2019}
\footnote{\url{http://argonaut.skymaps.info/}} extend far enough. However, they have some drawbacks:
\citetalias{drimmel} has a lower angular resolution, while \citetalias{green2019} covers only part of the sky.
Yet, we check that the use of \citetalias{drimmel} or \citetalias{green2019} provides us with vertical gradients, 
which are the same as using \citetalias{2015ApJ...798...88M} within 0.01 mag kpc$^{-1}$.

The moving modes of the dereddened colours and absolute magnitudes, with a window of 400 stars along $|Z|$, as functions of $|Z|$, 
are presented in Fig.~\ref{verticalgradients}.
We derive the following vertical gradients (in mag kpc$^{-1}$) for $400<|Z|<1714$ pc:
\begin{equation}
\label{gradient1}
\Delta(G_\mathrm{BP}-G_\mathrm{RP})_0/\Delta|Z|=-0.03\pm0.01\,,
\end{equation}
\begin{equation}
\label{gradient2}
\Delta(G_\mathrm{RP}-W3)_0/\Delta|Z|=-0.03\pm0.01\,.
\end{equation}
\begin{equation}
\label{gradient3}
\Delta M_\mathrm{G_\mathrm{BP}}/\Delta|Z|=+0.01\pm0.01\,,
\end{equation}
\begin{equation}
\label{gradient4}
\Delta M_\mathrm{G_\mathrm{RP}}/\Delta|Z|=+0.04\pm0.01\,,
\end{equation}
\begin{equation}
\label{gradient5}
\Delta M_\mathrm{W3}/\Delta|Z|=+0.07\pm0.01\,,
\end{equation}
The uncertainties of these gradients are computed as standard deviations of the gradients calculated when the space under 
consideration is varied in the range $400<|Z|<600$ pc, $60<|b|<75\degr$.

The modes within $|Z|<400$ pc are shown in Fig.~\ref{verticalgradients} by grey symbols.
They allow us to suggest that their gradients within $|Z|<400$ pc are similar to those within $|Z|>400$ pc.

\subsection{The narrow vertical cylinder}
\label{cylinder}

\begin{table*}
 \centering
\def\baselinestretch{1}\normalsize\footnotesize
\caption[]{The modes of the observables in a narrow cylinder of $(X^2+Y^2)^{0.5}<80$~pc. 
The uncertainty of all the mode values is 0.015 mag.
}
\label{column80pc}
\begin{tabular}[c]{rrrrrrr}
\hline
\noalign{\smallskip}
 $Z$ range, pc & Median $Z$, pc & $mode(M_\mathrm{G_\mathrm{BP}}+A_\mathrm{G_\mathrm{BP}})$ & $mode(M_\mathrm{G_\mathrm{RP}}+A_\mathrm{G_\mathrm{RP}})$ & $mode(M_\mathrm{W3}+A_\mathrm{W3})$ & $mode(G_\mathrm{BP}-G_\mathrm{RP})$ & $mode(G_\mathrm{RP}-W3)$ \\
\hline
\noalign{\smallskip}
$-1724$ -- $-168$ & $-384$ & $1.13$ & $-0.09$ & $-1.62$ & $1.22$ & $1.53$ \\ 
$-168$ -- $+114$  & $-40$  & $0.91$ & $-0.23$ & $-1.67$ & $1.14$ & $1.44$ \\ 
$+114$ -- $+1723$ & $+304$ & $1.09$ & $-0.11$ & $-1.64$ & $1.20$ & $1.53$ \\ 
\hline
\end{tabular}
\end{table*}

We analyse the modes of the observables as functions of $Z$ in a narrow space cylinder of $(X^2+Y^2)^{0.5}<80$~pc.
This allows us to estimate the intrinsic colours and absolute magnitudes of the nearby clump giants.
Then we can compare them with the values extrapolated in the thin coordinate layer in Sect.~\ref{thinlayer}.
Also, we can compare the observables near the Sun with those at high $|Z|$, behind the whole dust half-layer.
This allows us to derive the extinctions and reddenings through the whole dust half-layer.

The radius of the cylinder is set so that near the Sun it is limited by the Local Bubble. 
A wider cylinder may be contaminated by reddened stars. The cylinder contains 1224 stars. 
It is enough to calculate rather precise modes of the observables for only three ranges of $Z$.
They are presented in Table~\ref{column80pc}. Each range contains 408 stars.
The asymmetry of the ranges w.r.t. $Z=0$ is due to an uneven spatial distribution of the stars along $Z$.
The differences between the values at $Z=-384$ and $Z=+304$~pc is due to an offset of the Sun w.r.t. 
the dust layer mid-plane: more dust is below than above the Sun.

The stars in the range $(X^2+Y^2)^{0.5}<80$, $-168<Z<+114$ pc can be considered as unreddened ones.
More precisely, any reasonable distribution of dust suggests their average reddening to be $E(B-V)<0.01$ mag.
This is negligible w.r.t. the uncertainties which we consider further.

\begin{table*}
 \centering
\def\baselinestretch{1}\normalsize\footnotesize
\caption[]{Our estimates of the giant clump dereddened colours and absolute magnitudes.
}
\label{zero}
\begin{tabular}[c]{rrrrrr}
\hline
\noalign{\smallskip}
 $|Z|$, pc & $mode(M_\mathrm{G_\mathrm{BP}})$ & $mode(M_\mathrm{G_\mathrm{RP}})$ & $mode(M_\mathrm{W3})$ & $mode[G_\mathrm{BP}-G_\mathrm{RP})_0]$ & $mode[G_\mathrm{RP}-W3)_0]$ \\
\hline
\noalign{\smallskip}
$0$    & $0.910\pm0.01$ & $-0.235\pm0.01$ & $-1.670\pm0.01$ & $1.145\pm0.01$ & $1.435\pm0.01$ \\ 
$1700$ & $0.927\pm0.02$ & $-0.167\pm0.02$ & $-1.551\pm0.02$ & $1.094\pm0.02$ & $1.384\pm0.02$ \\ 
\hline
\end{tabular}
\end{table*}

Table~\ref{zero} presents our estimates of the giant clump dereddened colours and absolute magnitudes for $Z=0$.
These are averaged estimates for $R=0$ from Table~\ref{thinlayer40pc} and for $Z=0$ from Table~\ref{column80pc}.
Note that these estimates are in perfect agreement with each other.
Their uncertainty 0.01 mag is based on the assumption that they are independent.
Also, Table~\ref{zero} contains our estimates for $|Z|=1700$ pc, which are based on the estimates for $Z=0$ and
vertical gradients (\ref{gradient1})--(\ref{gradient5}), which are calculated within $400<|Z|<1714$ pc.
The uncertainty $\pm0.02$ includes the uncertainty of the gradients added in quadrature to the uncertainty at $Z=0$.
We compare Table~\ref{zero} with the theoretical estimates in Sect.~\ref{iso}.

Any reasonable vertical distribution of dust suggests that $>80$ per cent of the cumulative reddening or extinction
to infinity should manifest in modes of the observables for the outer ranges of the vertical cylinder.
This leads to a negligible systematic underestimation $\Delta E(B-V)<0.012$ mag of the reddening across the whole dust half-layer, 
if this reddening is $E(B-V)<0.06$ mag.
Therefore, the ranges of the vertical cylinder are convenient to derive the differences of the observables between the nearby 
stars and the stars behind the dust layer.

Hence, we estimate the extinctions and reddenings across the whole dust half-layer as the differences between the values for 
$Z=-40$ and the average values for $Z=-384$ and 304 pc from Table~\ref{column80pc}.
The values for $Z=-384$ and 304 pc must be corrected for vertical gradients (\ref{gradient1})--(\ref{gradient5}) in the interval 
$\Delta|Z|=(|-384-(-40)|+|304-(-40)|)/2=344$ pc. The reddening and extinction estimates are as follows: 
\begin{eqnarray}
\label{cylagbp}
&A_\mathrm{G_\mathrm{BP}}=(1.13+1.09)/2-(0.91+0.01\cdot0.344)= \nonumber\\
&0.20\pm0.02\,,
\end{eqnarray}
\begin{eqnarray}
\label{cylagrp}
&A_\mathrm{G_\mathrm{RP}}=(-0.09-0.11)/2-(-0.23+0.04\cdot0.344)= \nonumber\\
&0.12\pm0.02\,,
\end{eqnarray}
\begin{eqnarray}
\label{cylaw3}
&A_\mathrm{W3}=(-1.62-1.64)/2-(-1.67+0.07\cdot0.344)= \nonumber\\
&0.016\pm0.02\,,
\end{eqnarray}
\begin{eqnarray}
\label{cylebprp}
&E(G_\mathrm{BP}-G_\mathrm{RP})=(1.22+1.20)/2-(1.14-0.03\cdot0.344)= \nonumber\\
&0.08\pm0.02\,,
\end{eqnarray}
\begin{eqnarray}
\label{cylerpw3}
&E(G_\mathrm{RP}-W3)=(1.53+1.53)/2-(1.44-0.03\cdot0.344)= \nonumber\\
&0.10\pm0.02\,.
\end{eqnarray}

Estimates (\ref{cylagbp}), (\ref{cylagrp}), (\ref{cylebprp}), and (\ref{cylerpw3}) correspond to $E(B-V)\approx0.06$ mag
by use of all the extinction laws by \citetalias{davenport2014} for high-latitudes, \citetalias{schlafly2016} and 
\citetalias{wang2019}.
Estimate (\ref{cylaw3}) corresponds to very different $E(B-V)$ by use of the different laws: 0.048, 0.082, and 0.127 mag with the 
laws by \citetalias{davenport2014} for high-latitudes, \citetalias{schlafly2016} and \citetalias{wang2019}, respectively.

Results (\ref{cylagbp})--(\ref{cylerpw3}) define our extinction law. It is consistent with all the laws mentioned above.
This is seen from the following direct comparison. Given $A_\mathrm{G_\mathrm{BP}}=0.2$, the extinction laws 
by \citetalias{davenport2014} for high-latitudes, \citetalias{schlafly2016} and \citetalias{wang2019} provide
$A_\mathrm{G_\mathrm{RP}}=0.12$, 0.12, 0.12 and $A_\mathrm{W3}=0.02$, 0.01, 0.01 mag, respectively,
which are in good agreement with estimates (\ref{cylagrp}) and (\ref{cylaw3}).
The $A_\mathrm{W3}$ estimate from \citetalias{davenport2014} for high-latitudes is in best agreement with our 
estimate.
Given $A_\mathrm{G_\mathrm{BP}}=0.2$, the extinction law by \citetalias{ccm89} with $R_\mathrm{V}=3.1$ provides
$A_\mathrm{G_\mathrm{RP}}=0.13$.

Estimates (\ref{cylagbp})--(\ref{cylerpw3}) provide the CER $1.25^{+0.75}_{-0.45}$ in good agreement with the CERs,
which are mentioned in Sect.~\ref{thinlayer} for the other laws.

Estimates (\ref{cylagbp})--(\ref{cylerpw3}) are important, being independent of any model and based only on the differences in 
the observables between the nearby giants and those behind the dust layer.
We suppose that there is no other interpretation of these differences, except for the extinction/reddening across the whole dust 
half-layer. These estimates will be compared with some other estimates in Sect.~\ref{discuss}.

Our estimate (\ref{cylaw3}) of the IR extinction is quite uncertain. 
However, in combination with nearly zero IR extinction (\ref{thinaw3}) in the thin layer, this does not contradict
the variation of the extinction law with $|Z|$ or latitude, which has been found by \citetalias{davenport2014} and 
\citet{rv, g2013, g2016}.
These studies use a multiband photometry for stars up to $|Z|=25$ kpc.
They reveal a similar increase of the infrared-to-optical extinction ratios with $|Z|$ or latitude (see figure 9 in \citealt{g2016}).
This needs to be further explored, based on some more accurate IR extinction estimates.

\section{Discussion}
\label{theor}

\subsection{Absolute magnitudes}
\label{mw3}

Both $(G_\mathrm{BP}-G_\mathrm{RP})_0$ and $M_\mathrm{G_\mathrm{RP}}$ show an increase with age and [Fe/H], as seen in 
Fig.~\ref{parsec}, \ref{mist}, and \ref{iacbasti}.
This leads to a dependence of $M_\mathrm{G_\mathrm{RP}}$ on this colour.
Such a dependence of clump absolute magnitudes on its colours in the optical range is well known \citep{girardi2016}.
This makes it difficult to use the clump as a `standard candle' in the optical range.

In contrast, $(G_\mathrm{RP}-W3)_0$ increases mostly with [Fe/H], while $M_\mathrm{W3}$ -- mostly with age, as seen in 
Fig.~\ref{parsec} and \ref{mist}.  Hence, $M_\mathrm{W3}$ is nearly independent of $(G_\mathrm{RP}-W3)_0$.
This must be more or less true for all IR absolute magnitudes and colours of the clump.
Therefore, the IR absolute magnitudes of the clump seem to be suitable for using the clump as a `standard candle'.
However, Fig.~\ref{parsec} and \ref{mist} show a strong dependence of $M_\mathrm{W3}$ on age.
Unfortunately, age is a poorly known quantity, especially, with its variations with $|Z|$, $X'$ and from one galaxy to another.
Yet, to use the clump as a `standard candle' in the IR range, one must take into account a relation between age and absolute 
magnitude of the clump (see the discussion in \citealt{chen2017}).

Ignoring such a relation may be a reason for the inconsistency of the empirical estimates of the IR absolute magnitudes.
The most recent and precise estimates of $M_\mathrm{W3}$ near the Sun are as follows: \\
$-1.552\pm0.020$; \\
$-1.585\pm0.019$; \\
$-1.606\pm0.024$ obtained by \citet{yaz2013} for the fractional parallax $<0.05$, $<0.10$, and $<0.15$, respectively; \\
$-1.638\pm0.011$ \citepalias{ruiz2018} (we discuss this result in Sect.~\ref{discuss}); \\
$-1.676\pm0.028$ obtained by \citet{yaz2013} via the transformation from the 2MASS to the WISE photometric system; \\
$-1.752\pm0.068$ \citep{chen2017}; \\
$-1.67\pm0.02$ \citep{hawkins2017}; \\
$-1.66\pm0.02$ \citep{g2017}; \\
and our estimate \\
$-1.67\pm0.01$. \\
Note that the latter is based on the most complete sample.

The first four estimates deviate from each other and from the remaining estimates. 
The last five estimates agree with each other and with the value $-1.68$ at a level of $1\sigma$.
However, this agreement may be illusive, since some of these studies seem to ignore the dependences of $M_\mathrm{W3}$ on age and 
$|Z|$.

We put these estimates on the theoretical diagrams in Fig.~\ref{parsec} and \ref{mist}.
We find unreliable estimates of age and/or [Fe/H] for the first three estimates of $M_\mathrm{W3}$ by use of both PARSEC and MIST.
The reason of the deviation of the first three results is unknown.

The deviation of the result of \citetalias{ruiz2018} can be explained by a rather high $|Z|$ of their sample.
The median $|Z|$, mean $|Z|$, and standard deviation of $|Z|$ for their sample are 208, 206 and 85 pc, respectively.
\citetalias{ruiz2018} assume $A_\mathrm{W3}=0$ for their sample and, consequently, $M_\mathrm{W3}+A_\mathrm{W3}=M_\mathrm{W3}=-1.638$ 
mag.
The same $M_\mathrm{W3}+A_\mathrm{W3}\approx-1.64$ mag can be obtained from our estimates $M_\mathrm{W3}=-1.67$, (\ref{gradient5}) 
and (\ref{cylaw3}) followed by the median $M_\mathrm{W3}=-1.655$ and $A_\mathrm{W3}\approx0.013$.
Thus, the gradient of the absolute magnitude along $|Z|$ and the non-zero extinction can reconcile the estimate of 
\citetalias{ruiz2018} with the majority of the empirical estimates.

Among all the estimates of $M_\mathrm{W3}$, only the estimate of \citet{chen2017} provides a reliable age and [Fe/H] by use of MIST.
The inconsistency of the most empirical estimates of $M_\mathrm{W3}$ with MIST seems to indicate an imperfection of the MIST 
predictions rather than an error in the empirical estimates. 
The estimate of \citet{chen2017} can be closer to the remaining empirical estimates.
\citet{chen2017} use the rather large extinction coefficient $A_\mathrm{W3}/E(B-V)=0.349$ to corrrect the result
for $A_\mathrm{W3}$.
For comparison, the extinction laws by \citetalias{davenport2014} for low latitudes, \citetalias{schlafly2016} and 
\citetalias{wang2019} provide $A_\mathrm{W3}/E(B-V)=0.339$, 0.195, and 0.126, respectively.
With the latter, \citet{chen2017} would obtain $M_\mathrm{W3}$ several hundredths of a magnitudes fainter,
i.e. in better agreement with the bulk of the above estimates.

\subsection{Our results versus theoretical isochrones}
\label{iso}

\begin{table*}
 \centering
\def\baselinestretch{1}\normalsize\small
\caption[]{Age, [Fe/H] and their vertical gradients derived for the pairs 
$(G_\mathrm{BP}-G_\mathrm{RP})_0$ -- $M_\mathrm{G_\mathrm{RP}})$ and $(G_\mathrm{RP}-W3)_0$ -- $M_\mathrm{W3}$ 
from Table~\ref{zero} by use of different models.
The secondary solutions are given in brackets.
}
\label{hrrelation}
\begin{tabular}[c]{lllrr}
\hline
\noalign{\smallskip}
Model  & Age / [Fe/H] & $|Z|$ & $(G_\mathrm{BP}-G_\mathrm{RP})_0$ versus $M_\mathrm{G_\mathrm{RP}}$ & $(G_\mathrm{RP}-W3)_0$ versus $M_\mathrm{W3}$ \\
\hline
\noalign{\smallskip}
          &        & $Z=0$         & $2.2^{+0.3}_{0.2}$ ($4.6^{+0.6}_{-0.5}$) Gyr              & $2.5^{+0.3}_{-0.3}$ ($4.7^{+0.6}_{-0.5}$) Gyr \\
PARSEC    & Age    & $|Z|=1700$ pc & $8.3^{+1.0}_{-0.9}$ Gyr                                   & $7.8^{+1.0}_{-0.8}$ Gyr \\
          &        & gradient      & $3.6^{+1.0}_{-0.9}$ ($2.2^{+1.2}_{-1.0}$) Gyr kpc$^{-1}$  & $3.1^{+1.0}_{-0.9}$ ($1.8^{+1.1}_{-1.0}$) Gyr kpc$^{-1}$ \\
\hline
\noalign{\smallskip}
          &        & $Z=0$         & $-0.05\pm0.06$ ($-0.16\pm0.06$)                & $-0.05\pm0.06$ ($-0.19\pm0.06$) \\
PARSEC    & [Fe/H] & $|Z|=1700$ pc & $-0.32\pm0.06$                                 & $-0.33\pm0.06$ \\
          &        & gradient      & $-0.16\pm0.08$ ($-0.09\pm0.08$) dex kpc$^{-1}$ & $-0.16\pm0.08$ ($-0.08\pm0.08$) dex kpc$^{-1}$ \\
\hline
\noalign{\smallskip}
          &        & $Z=0$         & $1.9^{+0.2}_{-0.2}$ ($4.3^{+0.5}_{-0.5}$) Gyr  & $1.5^{+0.2}_{-0.2}$ ($6.5^{+0.8}_{-0.7}$) Gyr \\
MIST      & Age    & $|Z|=1700$ pc & $9.4^{+1.1}_{-1.0}$       Gyr                  & $10.3^{+1.3}_{-1.1}$ Gyr \\
          &        & gradient      & $4.4^{+1.2}_{-1.0}$ ($3.0^{+1.3}_{-1.1}$) Gyr kpc$^{-1}$ & $5.2^{+1.3}_{-1.1}$ ($2.2^{+1.5}_{-1.3}$) Gyr kpc$^{-1}$ \\
\hline
\noalign{\smallskip}
          &        & $Z=0$         & $-0.03\pm0.06$ ($-0.18\pm0.06$)                & $-0.14\pm0.06$ ($-0.39\pm0.06$) \\
MIST      & [Fe/H] & $|Z|=1700$ pc & $-0.40\pm0.06$                                 & $-0.52\pm0.06$ \\
          &        & gradient      & $-0.22\pm0.08$ ($-0.13\pm0.08$) dex kpc$^{-1}$ & $-0.22\pm0.08$ ($-0.08\pm0.08$) dex kpc$^{-1}$ \\
\hline
\noalign{\smallskip}
          &        & $Z=0$         & $2.8^{+0.3}_{-0.3}$ Gyr &  \\
IAC-BaSTI & Age    & $|Z|=1700$ pc & $5.6^{+0.7}_{-0.6}$ Gyr & \\
          &        & gradient      & $1.6^{+0.8}_{-0.7}$ Gyr kpc$^{-1}$ & \\
\hline
\noalign{\smallskip}
          &        & $Z=0$         & $-0.16\pm0.06$ &  \\
IAC-BaSTI & [Fe/H] & $|Z|=1700$ pc & $-0.31\pm0.06$ & \\
          &        & gradient      & $-0.09\pm0.08$ dex kpc$^{-1}$ & \\
\hline
\end{tabular}
\end{table*}

Table~\ref{zero} provides the empirical positions of the giant clump in the HR diagrams
`$(G_\mathrm{BP}-G_\mathrm{RP})_0$ versus $M_\mathrm{G_\mathrm{RP}}$' and 
`$(G_\mathrm{RP}-W3)_0$ versus $M_\mathrm{W3}$'.
We convert these positions into age and metallicity of the clump by use of the theoretical relations 
in Fig.~\ref{parsec}, \ref{mist} and \ref{iacbasti}.
The positions for $Z=0$ and $|Z|=1700$ pc are shown in these figures by the white crosses and stars, respectively.
The derived solutions, i.e. the pairs of the age and [Fe/H], are shown in Table~\ref{hrrelation}.
Also, Table~\ref{hrrelation} presents the vertical gradients of the age and [Fe/H].
They are calculated from the differences of the age and [Fe/H] for $Z=0$ and $|Z|=1700$ pc.

The uncertainties of the derived $\log age$ and [Fe/H] are $\pm0.05$ and $\pm0.06$ dex, respectively.
To calculate them, we take into account the uncertainties of the dereddened colours and absolute magnitudes from Table~\ref{zero}.

In some cases we find two solutions for $Z=0$. They are shown by two crosses in one plot.
Consequently, two gradients are calculated in such cases.
The secondary solutions, i.e. those with a higher age and lower [Fe/H], are given in Table~\ref{hrrelation} in brackets.
In the framework of the generally accepted theory of the giant clump \citep{girardi2016}, the secondary solutions
seem to be less reliable w.r.t. the primary ones.
Moreover, the secondary solutions are accompanied by the unrealistically low metallicities, as seen in Table~\ref{hrrelation}.

Table~\ref{hrrelation} shows an acceptable agreement of our results, except for the MIST predictions for $(G_\mathrm{RP}-W3)_0$ 
versus $M_\mathrm{W3}$.
The primary and secondary MIST age solutions for $Z=0$ and $(G_\mathrm{RP}-W3)_0$ versus $M_\mathrm{W3}$ (1.5 and 6.5 Gyr) 
are out of the reliable age range ($1.6-6$ Gyr) defined in Sect.~\ref{thinlayer}.
Moreover, the MIST isochrone for 1.5 Gyr is a typical isochrone of high-mass clump giants. 
Consequently, MIST claims that high-mass clump giants dominate in our sample for $Z=0$. This seems unlikely, as discussed
in Sect.~\ref{thinlayer}.
The most important here is the discrepancy between the MIST estimates for $(G_\mathrm{BP}-G_\mathrm{RP})_0$ versus 
$M_\mathrm{G_\mathrm{RP}}$ and $(G_\mathrm{RP}-W3)_0$ versus $M_\mathrm{W3}$. 
This may indicate an issue in the MIST predictions for $W3$.
Most likely, this is an issue in the colour--$T_\mathrm{eff}$ relation and/or bolometric correction used.

We average the primary estimates for $(G_\mathrm{BP}-G_\mathrm{RP})_0$ versus $M_\mathrm{G_\mathrm{RP}}$ and conclude with 
caution that age and [Fe/H] of the clump change from $2.3\pm0.5$ to $7.8\pm2.2$ Gyr and from $-0.08\pm0.08$ to $-0.34\pm0.06$, 
respectively, when $|Z|$ changes from 0 to 1700 pc.
Consequently, the vertical gradients of the age and [Fe/H] are $3.2\pm1.6$ Gyr kpc$^{-1}$ and $-0.16\pm0.07$ dex kpc$^{-1}$, 
respectively.

Our age gradient agrees with an estimate of 4 Gyr kpc$^{-1}$ by \citet{casagrande2016}.
Moreover, the vertical gradients of $M_\mathrm{W3}$ and age along $|Z|$ are combined into the vertical gradient of $M_\mathrm{W3}$ 
along age 0.02$^{+0.03}_{-0.01}$ mag Gyr$^{-1}$.
This agrees with its estimate $0.017\pm0.004$ mag Gyr$^{-1}$ by \citet{chen2017}.

Our [Fe/H] gradient is in the middle of the diversity of its estimates presented, for example, by \citet{onaltas2016}.
In particular, our [Fe/H] gradient agrees with the estimate $-0.15\pm0.01$ dex kpc$^{-1}$ by \citet{huang2015} for a large sample 
of clump giants within $-1<X'<1$ and $|Z|<3$ kpc.
However, our [Fe/H] gradient disagrees with the estimate $-0.31\pm0.03$ dex kpc$^{-1}$ by \citet{soubiran2008} for a small sample 
of local clump giants.

Our results can be applied to some contradictory cases.
\citetalias{ruiz2018} attempt to create a sample of clump giants with a very low reddening of $E(B-V)<0.015$ mag. 
They select stars with reddening estimates from the 2D reddening map of \citetalias{sfd} and other maps with very low estimates of 
reddening far from the Galactic mid-plane. 
\citetalias{ruiz2018} obtain the median $E(B-V)=0.01$ mag for their sample. The majority of their giants are located at high $|Z|$.
For their median $|Z|=208$ pc, our results by use of PARSEC suggest a median age of about 3 Gyr and the median 
$\mathrm{[Fe/H]}=-0.08$.
\citetalias{ruiz2018} claim that the position of their sample in the HR diagram with the $G-K_s$ colour `seems to nicely fit' a 
PARSEC isochrone.   However, their figure 8 shows that the location of the clump better fits an isochrone of 5 Gyr and 
$\mathrm{[Fe/H]}=0$.
This PARSEC isochrone is $\Delta(G-K_s)=0.1$ mag redder than the isochrone for 3 Gyr and $\mathrm{[Fe/H]}=-0.08$.
Thus, there are two competing sets of the parameters to fit the observed clump by a PARSEC isochrone: \citetalias{ruiz2018}'s 5 Gyr, 
$\mathrm{[Fe/H]}=0$, $E(B-V)=0.01$ mag versus our 3 Gyr, $\mathrm{[Fe/H]}=-0.08$ and a much higher reddening.

To estimate the latter, we must accept the isochrone colour difference $\Delta(G-K_s)=0.1$ mag as the difference 
$\Delta E(G-K_s)=0.1$ mag between the reddening estimates.
It is converted to $\Delta E(B-V)=0.044$ by use of the extinction law by \citetalias{wang2019}.
Hence, the needed, much higher reddening is $E(B-V)=0.01+0.044=0.054$ mag.
This is in perfect agreement with our estimate of the median reddening at $|Z|=208$ pc.
Finally, our set of the parameters, i.e. 3 Gyr, $\mathrm{[Fe/H]}=-0.08$, and $E(B-V)=0.054$ mag, better corresponds to the 
PARSEC's description of the giant clump than the set of \citetalias{ruiz2018}.
In this case, only a few giants in the sample of \citetalias{ruiz2018} actually have $E(B-V)<0.015$~mag.
This may be a typical case when the reddening far from the Galactic mid-plane is underestimated because slightly reddened stars 
are erroneously considered as unreddened.
Also, this case demonstrates that one should not consider stars as unreddened, based only on their low reddening
estimates in a map, without a detailed investigation of this issue.

\subsection{Age--metallicity relation}
\label{amrsection}

\begin{figure}
\includegraphics{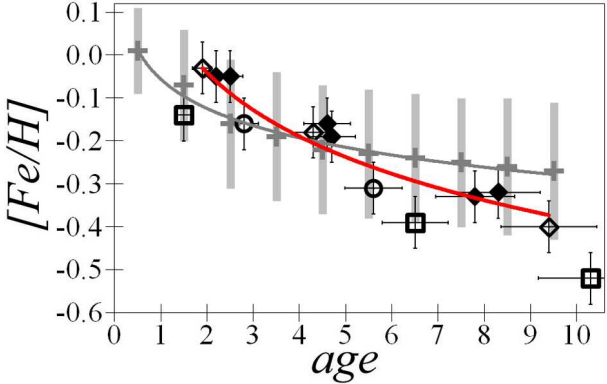}
\caption{AMR by use of our results from Table~\ref{hrrelation}:
the PARSEC estimates -- black filled diamonds,
the MIST estimates for $(G_\mathrm{BP}-G_\mathrm{RP})_0$ versus $M_\mathrm{G_\mathrm{RP}}$ -- black open diamonds,
the MIST estimates for $(G_\mathrm{RP}-W3)_0$ versus $M_\mathrm{W3}$ -- the black open squares, and 
the IAC-BaSTI estimates -- the black open circles.
The approximation of the estimates for $(G_\mathrm{BP}-G_\mathrm{RP})_0$ versus $M_\mathrm{G_\mathrm{RP}}$ -- 
the red curve.
AMR from \citet{soubiran2008} -- the grey crosses with the wide error bars and the grey logarithmic approximating curve.
}
\label{amr}
\end{figure}

We can derive the AMR in the vertical space cylinder.
All our estimates from Table~\ref{hrrelation} are shown in Fig.~\ref{amr} by different black symbols.
The MIST estimates for $(G_\mathrm{RP}-W3)_0$ versus $M_\mathrm{W3}$ (the open squares),
the IAC-BaSTI estimates (the open circles) and the remaining estimates show slightly different trends:
$\mathrm{[Fe/H]}=-0.08-4.4\times10^{-11}\cdot age$, 
$\mathrm{[Fe/H]}=-0.01-5.4\times10^{-11}\cdot age$, and
$\mathrm{[Fe/H]}=+0.05-4.8\times10^{-11}\cdot age$, respectively.
The former seems to be unreliable, since it predicts too low metallicity for newborn stars.
Thus, the MIST estimates for $(G_\mathrm{RP}-W3)_0$ versus $M_\mathrm{W3}$ appear outliers in Fig.~\ref{amr},
as in Table~\ref{hrrelation}.
The remaining estimates show good agreement with each other, as for the different models, as for the primary 
or secondary solutions.  It is worth noting that our estimates are model dependent.
Hence, this agreement shows a convergence of the models for the optical bands.
For the $(G_\mathrm{BP}-G_\mathrm{RP})_0$ versus $M_\mathrm{G_\mathrm{RP}}$ estimates the logarithmic approximation
$\mathrm{[Fe/H]}=4.56-0.495\log age$ is the most reliable. It is shown by the red curve.
As seen from Fig.~\ref{amr}, it satisfies all the estimates within their uncertainties, except 
the MIST estimates for $(G_\mathrm{RP}-W3)_0$ versus $M_\mathrm{W3}$.

For comparison, the AMR from \citet{soubiran2008} for a smaller sample of local clump giants is shown in Fig.~\ref{amr} by the 
grey crosses with the wide error bars.   Its logarithmic approximation is shown by the grey curve.
It is seen that our results are more precise due to a larger sample.
Our AMR is steeper, but consistent with that of \citet{soubiran2008} within the uncertainties.

\begin{table*}
 \centering
\def\baselinestretch{1}\normalsize\normalsize
\caption[]{The estimates of the extinction or reddening across the whole dust half-layer below or above the Sun.
}
\label{summ}
\begin{tabular}[c]{lcc}
\hline
\noalign{\smallskip}
 Model / map / result      & Original estimate & $E(B-V)$ by \citetalias{davenport2014} \\
\hline
\noalign{\smallskip}
This study, narrow cylinder      & $A_\mathrm{G_\mathrm{BP}}=0.20$            & 0.060 \\
This study, narrow cylinder      & $A_\mathrm{G_\mathrm{RP}}=0.12$            & 0.060 \\
This study, narrow cylinder      & $A_\mathrm{W3}=0.016$                      & 0.048 \\
This study, narrow cylinder      & $E(G_\mathrm{BP}-G_\mathrm{RP})=0.08$      & 0.060 \\
This study, narrow cylinder      & $E(G_\mathrm{RP}-W3)=0.10$                 & 0.061 \\
Model \citetalias{arenou}        & $A_\mathrm{V}=0.100$                       & 0.031 \\ 
Model \citetalias{drimmel}       & $E(B-V)=0.018$                             & 0.018 \\
Model \citetalias{av}            & $A_\mathrm{V}=0.214$                       & 0.066 \\
2D map \citetalias{sfd}                 & $E(B-V)=0.018$                      & 0.018 \\
2D map \citetalias{2015ApJ...798...88M} & $E(B-V)=0.015$                      & 0.015 \\
3D map \citetalias{g17}                 & $E(J-K_s)=0.03$                     & 0.060 \\
3D map \citetalias{lallement2019}       & $E(B-V)=0.010$                      & 0.010  \\
3D map \citetalias{green2019}           & $E(B-V)=0.051$                      & 0.051 \\ 
\hline
\end{tabular}
\end{table*}


\section{Reddening across the whole dust layer}
\label{discuss}

Table~\ref{summ} summarizes some estimates of the reddening across the whole dust half-layer below or above the Sun.
In perfect agreement with the estimate of \citet{gm2018}, mentioned in Sect.~\ref{intro}, the estimates from this study converge 
to $E(B-V)=0.06$ mag\footnote{All mentioned extinction laws provide similar estimates in the optical range.}.
The only exception is $A_\mathrm{W3}$.
However, it is rather uncertain, and its conversion to $E(B-V)$ strongly depends on an extinction law applied.
The extinction law by \citetalias{davenport2014} for high-latitudes appears to be the most reliable in this case.

We do not provide uncertainties on the estimates in Table~\ref{summ}, since some of the estimates
(i) are upper or lower limits without any uncertainty estimate (e.g. \citetalias{arenou}
\footnote{\citetalias{arenou} apply the constraint $A_\mathrm{V}<0.1$~mag for $|b|\ge60\degr$.}),
(ii) have very different estimates of their uncertainties from different studies (e.g. \citetalias{sfd}), or
(iii) are, apparently, underestimated or overestimated.
Anyway, the formal uncertainties of the estimates in Table~\ref{summ} make little sense, since they cannot explain the large 
diversity of the estimates. In particular, Table~\ref{summ} shows several widely used estimates of $E(B-V)\ll0.06$,
in contrast to our estimates.

\citet{gm2018} suggest some reasons of possible underestimation of the reddening across the whole dust half-layer in \citetalias{sfd} 
and \citetalias{2015ApJ...798...88M}.
Moreover, there is a lot of previous findings of this underestimation (a review is given by \citealt{astroph}).
For example, \citet{berry2012} note:
`It is confirmed beyond doubt that there are some systematic problems with the normalization\footnote{This normalization is
the reddening-to-emission calibration, which we discuss later in this Section.}
of SFD extinction map'. 
\citetalias{green2019} note:
`Both this work and \citet{green2018} have identified systematic trends in reddening when comparing with the Planck14 
and SFD dust maps, particularly for reddenings of $E(g-r)\lesssim0.1$ mag. These systematics could be due to the 
uncertain zero-point of the dust reddening (e.g., what is the absolute reddening of some reference point on the sky?) 
and variation in $R_\mathrm{V}$ across the high-Galactic-latitude sky.'

The uncertainty of high reddenings in \citetalias{sfd} is 16 per cent. 
However, it is much worse for $E(B-V)\lesssim0.1$ mag, as admitted by the authors. 
This uncertainty can be estimated from the standard deviation of the reddening-to-emission calibration residuals as $\pm0.028$ mag.
Moreover, by use of {\it counts} of galaxies at high-latitudes, \citetalias{sfd} obtain approximately twice larger 
reddening-to-emission calibration coefficient than that obtained by use of the {\it reddened colours} of these galaxies.
The latter coefficient is finally adopted in \citetalias{sfd}.
This may lead to a systematic underestimation of the reddening across the whole dust layer by use of the adopted calibration 
at a level of a few hundredths of a magnitude.

\citetalias{2015ApJ...798...88M} may also suffer from some reddening-to-emission calibration errors.
In their figure 11, bottom left, \citetalias{2015ApJ...798...88M} show a systematic trend in the calibration 
as a function of hot dust temperature.
This trend suggests at least $\pm0.02$ mag systematic uncertainty of the calibration.
This may lead to a similar systematic underestimation of the reddenings.

The similarity of the \citetalias{drimmel} and \citetalias{sfd} estimates in Table~\ref{summ} is defined by the method of the 
former: the \citetalias{drimmel}'s `rescaling factors for latitudes $|b|>30\degr$ are based on the \citetalias{sfd} Galactic 
extinction map' \citepalias{drimmel}.

Thus, \citetalias{lallement2019} seems to be the only recent all-sky 3D reddening map with very low estimates of the reddening 
across the whole dust half-layer.
Possible systematic errors of \citetalias{lallement2019} due to some biases have been discussed by \citet{polarization}.

\citet{teerikorpi1990} notes on such biases:
`a large part, maybe all, of the discrepancy between stellar reddenings and other high-latitude extinction indicators
can be ascribed to the reddening bias'.   Two kinds of such biases are analysed by \citet{teerikorpi1990}.
(i) Due to the Malmquist effect, distant stars in a magnitude limited sample have a bluer average intrinsic colour than such colour 
of similar stars in a local distance limited sample.
(ii) Distant, highly reddened stars tend to remain below the detection limit of a magnitude limited sample.
This leads to an underrepresentation of highly reddened stars in such a sample.
Hence, its distant stars have a bluer average reddened colour.
In turn, such blueward biases of a colour of distant stars lead to an underestimation of the reddening as the difference
between the colour of nearby (unreddened) and distant (reddened) stars.

After an analysis of these biases, \citet{teerikorpi1990} suggests that `the average reddening reaches $E(B-V)=0.04$ mag at 
400 pc above the Galactic plane' \citepalias{sfd}.

Both biases are common in reddening maps and models. For example, a previous (albeit rather similar) version of 
\citetalias{lallement2019} provided by \citet[][hereafter LVV14]{lallement2014} admits the bias (ii) in their results:
`There is a limitation in the brightness of the target stars, and the subsequent lack of strongly reddened 
stars results\ldots There is for the same reasons a bias towards low opacities\ldots'.

\citet{polarization} discuss another bias -- (iii) an overestimation of the intrinsic colours due to the difficulties in the 
selection of unreddened stars.
In other words, this bias originates from a use of slightly reddened stars as unreddened.

Note that these biases also may lead to the above mentioned reddening-to-emission calibration systematic errors in
\citetalias{sfd} and \citetalias{2015ApJ...798...88M} due to underestimation of reddening of stars, quasars and galaxies 
used in the calibration.

These biases have been eliminated to some extent in the three consecutive versions of the GSF map:    
\citet[][hereafter GSF15]{green2015}, \citet[][hereafter GSF18]{green2018}, and \citetalias{green2019}.
\citetalias{green2018} have recognized the underestimation of low reddenings due to an offset in the colour zero-point 
of the stellar templates.
Therefore, in this sequence of the map versions the new templates are bluer than the old ones.
Also, \citetalias{green2018} has realized that \citetalias{green2015} 
`tends to infer zero reddening when the true reddening is below a few hundredths of a magnitude.'
As a result, \citetalias{green2019} provides higher reddenings at high-latitudes than those from \citetalias{green2015}.

\subsection{Reddening of Galactic globular clusters}

\begin{figure}
\includegraphics{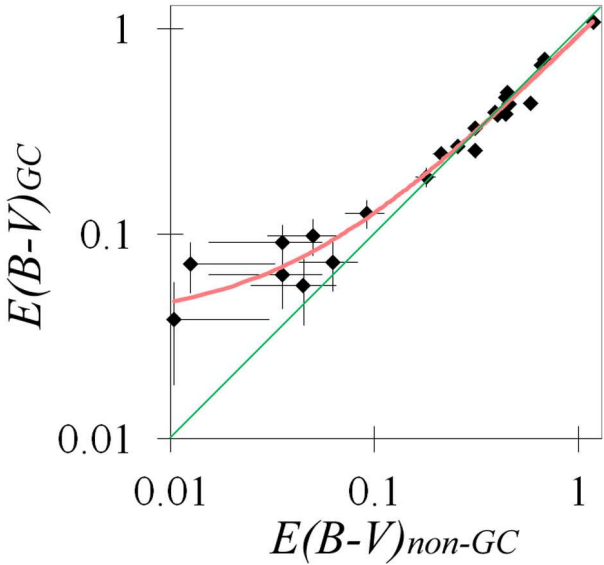}
\caption{$E(B-V)$ from \citetalias{green2019} in the non-GC versus GC voxels for the 23 GCs.
The linear trend is shown by the thick red line, while the one-to-one relation -- by the thin green line.
}
\label{gc}
\end{figure}

Direct measurements of the extinction and reddening of objects far from the Galactic mid-plane may be an ultimate solution of 
this issue. However, such measurements are rare.
Examples are given by \citet{ngc5904, ngc6205} for stars of the Galactic globular clusters (GCs) NGC\,5904 (M5) and NGC6205\, (M13) 
at the latitudes of about $+47\degr$ and $+41\degr$, respectively.
For both the clusters, the optical extinction is found to be twice as high as generally accepted.
Thus, high-latitude GCs seem to be promising probes of the reddening across the whole dust layer.

\citetalias{green2019} is based on {\it Gaia} DR2 parallaxes and a deep multiband photometry from 2MASS and the Panoramic Survey 
Telescope and Rapid Response System Data Release 1 (Pan-STARRS DR1, \citealt{chambers2016}).
This makes \citetalias{green2019} the first 3D reddening map with accurate reddening estimates far beyond the dust layer 
at high-latitudes.

\citetalias{green2019} use thousands GC stars in a typical voxel (space cell) within a GC field to derive their reddening,
distance modulus, absolute magnitude, and metallicity.
These results agree with those for the same GCs obtained by \citet{bernard2014} and \citet{ngc5904, ngc6205} from the
Pan-STARRS photometry of the same GC stars by another method.
The most important that the \citetalias{green2019}'s rather high reddening estimates for these GCs agree with the estimates by
\citet{ngc5904, ngc6205}.

In contrast to the voxels within the GC fields with thousands stars, \citetalias{green2019} use only a hundred stars in a typical 
voxel outside the GC fields, far from the Galactic mid-plane.
Therefore, the \citetalias{green2019}'s reddening estimates in the GC voxels seem to be more reliable than those in the 
surrounding non-GC voxels.

We select 23 GCs, whose stars dominate in the corresponding voxels of \citetalias{green2019}.
This dominance is evident in colour--magnitude diagrams.
Typically, such a diagram contains an order of 10\,000 stars for each GC with an accurate photometry.
These GCs have a distance $<10.5$ kpc from the Sun 
(taken from the data base of GCs by \citealt{harris}\footnote{\url{https://www.physics.mcmaster.ca/~harris/mwgc.dat}}, 2010 revision)
and an angular diameter $\ge10$ arcmin (taken from \citealt{bica2019}).
The latter provides, at least, five adjacent voxels where GC stars dominate.

For each GC, Fig.~\ref{gc} compares an average \citetalias{green2019} $E(B-V)$\footnote{Based on the \citetalias{green2019} 
extinction law, we adopt $E(B-V)=1.04E(g_\mathrm{P1}-r_\mathrm{P1})$, where $g_\mathrm{P1}$ and $r_\mathrm{P1}$ are the Pan-STARRS
filters.} 
estimate in the voxels with a dominance of the GC stars and in surrounding non-GC voxels with virtually no such stars, i.e. those 
between 2 and 3 radii of the GC.
The linear trend
\begin{equation}
\label{gsf}
E(B-V)_{GC}=0.038+0.893\,E(B-V)_{non-GC}
\end{equation}
with a linear correlation coefficient of 0.99 is shown in Fig.~\ref{gc} by the thick red curve.
The mean \citetalias{green2019} reddening in all non-GC voxels near the Galactic poles, far from the Galactic mid-plane, 
is $E(B-V)=0.015$ mag.
Consequently, given equation (\ref{gsf}), this provides a \citetalias{green2019} estimate of the mean reddening across
the whole dust half-layer below or above the Sun, derived for the GC stars: $E(B-V)=0.038+0.893\cdot0.015=0.051$ mag.
This value, presented in Table~\ref{summ}, supports our estimates.

The difference between GC and non-GC voxels, shown in Fig.~\ref{gc} by the thick red curve, is not due to a dust inside 
the GCs, since this would suggest the presence of such dust only inside the high-latitude GCs.
This difference may suggest an underestimation of the reddening across the whole dust half-layer by the conventional approaches.
This needs futher investigation.

\section{Conclusions}
\label{conclusions}

{\it Gaia} DR2 provides first accurate parallaxes and optical photometry for a sample of giants in the clump domain of 
the HR diagram, which is complete for a space across the whole Galactic dust layer near the Sun.
In this paper we have used the observables 
$G_\mathrm{BP}-G_\mathrm{RP}$, 
$G_\mathrm{RP}-W3$, 
$M_\mathrm{G_\mathrm{BP}}+A_\mathrm{G_\mathrm{BP}}=G_\mathrm{BP}+5+5\,\log_{10}\varpi$,
$M_\mathrm{G_\mathrm{RP}}+A_\mathrm{G_\mathrm{RP}}=G_\mathrm{RP}+5+5\,\log_{10}\varpi$, and
$M_\mathrm{W3}+A_\mathrm{W3}=W3+5+5\,\log_{10}\varpi$.
They are based on the {\it Gaia} DR2 parallaxes $\varpi$, photometry in the $G_\mathrm{BP}$, $G_\mathrm{RP}$ bands
and {\it WISE} photometry in the $W3$ band.
We created a complete sample of 101\,810 giants in a space cylinder with the radius 700~pc 
around the Sun, up to $|Z|<1800$~pc along the $Z$ Galactic coordinate.

We assumed that the spatial variations of the modes of the observables reflect the spatial variations of the extinction and 
reddening, in combination with some linear vertical gradients of the intrinsic colours and absolute magnitudes of the giant clump.

This approach allowed us to derive these characteristics of the clump in combination with the estimates of the extinction and 
reddening along the Galactic mid-plane and across the whole dust layer.
We derived the intrinsic colours and absolute magnitudes of the nearby clump giants in a thin coordinate layer and in a narrow 
vertical cylinder.
These two sets of the derived characteristics agree within 0.005 mag.
Also, we estimated the vertical gradients of the intrinsic colours and absolute magnitudes of the clump within $|Z|<1700$ pc 
by use of the observables in the space beyond the dust layer and the reddening estimates 
from the \citetalias{2015ApJ...798...88M} map.

Finally, for $|Z|<1.7$ kpc we found: \\
$M_\mathrm{G_\mathrm{BP}}=(0.91\pm0.01)+(0.01\pm0.01)\,|Z|$, \\
$M_\mathrm{G_\mathrm{RP}}=(-0.235\pm0.01)+(0.04\pm0.01)\,|Z|$, \\
$M_\mathrm{W3}=(-1.67\pm0.01)+(0.07\pm0.01)\,|Z|$, \\
$(G_\mathrm{BP}-G_\mathrm{RP})_0=(1.145\pm0.01)-(0.03\pm0.01)\,|Z|$ and \\
$(G_\mathrm{RP}-W3)_0=(1.435\pm0.01)-(0.03\pm0.01)\,|Z|$, \\
where $Z$ is expressed in kpc.
The obtained $M_\mathrm{W3}$ agrees with the recent literature estimates.
The other obtained values have no robust empirical estimates in the literature, since they are obtained for the first time at 
a precision level of 0.01 mag.

We compared the derived clump intrinsic colours and absolute magnitudes with the theoretical predictions from PARSEC, MIST and 
IAC-BaSTI.
This allowed us to estimate the clump's median age and [Fe/H] with their linear vertical gradients within $|Z|<1.7$ kpc 
as $(2.3\pm0.5)+(3.2\pm1.6)\,|Z|$ Gyr and $(-0.08\pm0.08)-(0.16\pm0.07)\,|Z|$, respectively, where $Z$ is expressed in kpc.
These results agree with the recent empirical and theoretical estimates, particularly, with those from the review 
of \citet{girardi2016}.
The predictions from the three models agree for the optical bands.
However, the PARSEC and MIST predictions disagree for $W3$ (IAC-BaSTI provides no information for $W3$).
This may indicate an issue in the MIST colour--$T_\mathrm{eff}$ relation and/or bolometric correction for $W3$.

Serendipitously, all the models give similar age--metallicity relations by use of our results in the optical range.
This similarity suggests that the models converge to a realistic representation of nature.
For the `$(G_\mathrm{BP}-G_\mathrm{RP})_0$ versus $M_\mathrm{G_\mathrm{RP}}$' estimates the logarithmic approximation
$\mathrm{[Fe/H]}=4.56-0.495\log age$ is the most reliable.

The obtained estimates of the extinction and reddening across the whole dust half-layer below or above the Sun converge to the 
mean reddening $E(B-V)=0.06$ mag. This agrees with the recent estimates from the 3D reddening map of \citetalias{green2019} 
for high-latitude Galactic globular clusters, the \citetalias{g17} map, the \citetalias{av} model, and some other studies.
Therefore, our estimate favours higher reddenings across the whole dust half-layer w.r.t. the values advocated by 
\citetalias{sfd}, \citetalias{2015ApJ...798...88M}, \citetalias{drimmel}, and \citetalias{lallement2019}.
A further investigation of this issue is important for a correct estimation of the extinction for high-latitude extragalactic objects.

\section*{Data availability}

The data underlying this article will be shared on reasonable request to the corresponding author.

\section*{Acknowledgements}

We thank the anonymous reviewers and Xiaodian Chen for useful comments, Aaron Dotter, Santi Cassisi and Leonid Petrov for 
discussion of our results, and Jeremy Mutter for assistance with English syntax.

The research described in this paper makes use of Filtergraph \citep{filtergraph}, an online data visualization tool developed 
at Vanderbilt University through the Vanderbilt Initiative in Data-intensive Astrophysics (VIDA) and the Frist Center for 
Autism and Innovation (FCAI, \url{https://filtergraph.com}).
The resources of the Centre de Donn\'ees astronomiques de Strasbourg, Strasbourg, France
(\url{http://cds.u-strasbg.fr}), including the SIMBAD database and the X-Match service, were widely used in this study.
This work has made use of data from the European Space Agency (ESA) mission {\it Gaia}
(\url{https://www.cosmos.esa.int/gaia}), processed by the {\it Gaia} Data Processing and Analysis Consortium
(DPAC, \url{https://www.cosmos.esa.int/web/gaia/dpac/consortium}).
This publication makes use of data products from the {\it Wide-field Infrared Survey Explorer}, which is a joint project
of the University of California, Los Angeles, and the Jet Propulsion Laboratory/California Institute of Technology.

\bsp	
\label{lastpage}
\end{document}